\documentclass[11pt]{article}
\usepackage{graphicx,amsmath,bm, amsthm,mathrsfs,amssymb, braket, verbatim}
\usepackage{caption}
\usepackage{subcaption}
\usepackage[usenames]{color}
\usepackage{ulem,mathtools}
\usepackage{pdfpages}
\usepackage{lscape}
\usepackage{authblk}
\usepackage{cite}
\usepackage{tcolorbox}
\usepackage[section]{placeins}
\usepackage{placeins}

\newcommand{\ee}{\end{equation}}

\newcommand{\reff}[1]{(\ref{#1})}
\newcommand{\beq}{\begin{equation}}
\newcommand{\eeq}[1]{\label{#1}\end{equation}}
\newcommand{\beqa}{\begin{eqnarray}}
\newcommand{\eea}{\end{eqnarray}}
\newcommand{\eeqa}[1]{\label{#1}\end{eqnarray}}
\newcommand{\beg}{\begin{equation*}}
\newcommand{\eeg}{\end{equation*}}

\newcommand{\m}{\!-\!}

\newcommand{\bsplit}{\begin{split}}
\newcommand{\esplit}{\end{split}}

\usepackage{circuitikz} 
\usepackage[capposition=bottom]{floatrow} 
\usepackage{epigraph} 
\usepackage{appendix}
\usepackage{rotating}
\allowdisplaybreaks

\title{Convergent perturbative series via finite path integral limits: application to energy at strong coupling \\ of the anharmonic oscillator}
\author[]{Ariel Edery\thanks{aedery@ubishops.ca}}
\affil[]{Department of Physics and Astronomy, Bishop's University, 2600 College Street, Sherbrooke, Qu\'{e}bec, Canada, J1M 1Z7.\vspace{1em}}
\begin{document}
\date{}
\maketitle
\begin{abstract}
Solving quantum field theories at strong coupling remains a challenging task. The main issue is that the usual perturbative series are asymptotic series which can be useful at weak coupling but break down completely at strong coupling. In this work, we show that  if the limits of integration in the path integral are finite, the perturbative series is remarkably an absolutely convergent series which works well at strong coupling. For now, we apply this perturbative approach to $\lambda \,\phi^4$ theory in $0+0$ dimensions (a basic integral) and $0+1$ dimensions (quartic anharmonic oscillator). As a further application, we also consider the sextic anharmonic oscillator. For the basic integral, we show that finite integral limits yields a convergent series whose values are in agreement with exact analytical results at any coupling. This worked even when the asymptotic series was not Borel summable. It is well known that the perturbative series expansion in powers of the coupling for the energy of the (quartic and sextic) anharmonic oscillator yields an asymptotic series and hence fails at strong coupling. In quantum mechanics, if one is interested in the energy, it is often easier to use Schr\"odinger's equation to develop a perturbative series than path integrals. Finite path integral limits are then equivalent to placing infinite walls at positions $-L$ and $L$ in the potential where $L$is positive, finite and can be arbitrarily large. With walls, the series expansion for the energy is now convergent and approaches the energy of the (quartic and sextic) anharmonic oscillator as the walls are moved further apart. We use the convergent series to calculate the ground state energy at weak, intermediate and strong coupling. At strong coupling, the result from the series agrees with the exact (correct) energy to within $0.1\%$, a remarkable result in light of the fact that at strong coupling the usual perturbative series diverges badly right from the start. 
\end{abstract}
\setcounter{page}{1}
\newpage
\section{Introduction}\label{Intro}
In quantum field theory (QFT), perturbative series in powers of the coupling have had remarkable success at weak coupling but are not useful (break down) at strong coupling. The reason is that the usual perturbative series is an asymptotic series and at strong coupling the series diverges immediately (i.e. starting at zeroth order, as the order increases, the series departs more and more from the correct answer and never has a value close to it at any order). At weak coupling the series at low orders improves as the order increases and settles/plateaus over a range of orders to a value that is close to the correct answer (for realistic theories, matches experiment to a high level of accuracy). The series does ultimately diverge at large orders. Dyson gave a famous argument as to why the perturbative series in powers of the coupling in QFT should yield an asymptotic series \cite{Dyson}. However, the divergence of the perturbative series raises an important question. The integrand in a path integral is an exponential and an exponential has an exact series expansion (it has an infinite radius of convergence). So why should the perturbative series based on the expansion of an exponential yield a divergence? Recall that each term in the series is obtained from integrals with infinite limits. As we will show, the infinite integral limits are what causes the problem for the perturbative series. The issue is resolved if one replaces the infinite limits by finite ones; one then obtains an absolutely convergent series which approaches the correct result as the finite limit is made sufficiently large. This is explained in more detail in section 2.1 below and also summarized in bullet form in the same section. In the conclusion we discuss why Dyson's argument no longer applies with the finite limits.       

The crucial point is that this perturbative approach based on finite path integral limits yields an absolutely convergent series that allows one to make reliable calculations and predictions at strong coupling. This should in principle hold true for realistic QFTs like QED and QCD in $3+1$ dimensions. Here, to highlight in a clear and transparent fashion the main features of this approach, we reduce technicalities by focusing on two simpler models. We apply the convergent series to $\lambda\,\phi^4$ theory in $0+0$ dimensions (a basic integral with quadratic and quartic terms) and $0+1$ dimensions (quartic anharmonic oscillator). As a further application, we also consider the sextic anharmonic oscillator. The integrand of the basic integral is $e^{-\frac{1}{2} a \,x^2 -\lambda \,x^4}$ with coupling $\lambda>0$. There are two cases:  $a>0$ and $a<0$. In both cases the original integral yields an exact analytical result but the usual perturbative series in powers of the coupling $\lambda$ is an asymptotic series. Using finite integral limits we obtain a convergent series in powers of the coupling whose values are in agreement with the exact analytical results. To make contact with field theory techniques, we also show that one can define a generating functional with \textit{finite limits} from which we can obtain the same convergent series by taking functional derivatives. The basic integral with $a<0$ yields an asymptotic series which is not Borel summable. A positive feature is that finite integral limits work as is (without any modification) for this non-Borel summable case.  

We then use our perturbative approach to obtain the ground state energy of the (quartic and sextic) anharmonic oscillator at strong coupling. It is well known that the usual perturbative series in powers of the coupling for the ground state energy of the anharmonic oscillator is an asymptotic series \cite{Wu1,Wu2,Bender1,Bender2,Marino,Strocchi}. Since we are interested in the energy it is easier to use Schr\"odinger's equation than path integrals to develop the perturbative series. For Schr\"odinger's equation, finite path integral limits are equivalent to placing infinite walls at positions $-L$ and $L$ in the potential (here $L$ is finite and positive but can be arbitrarily large). With walls, the energy series expansion is no longer an asymptotic series but an absolutely convergent series. The walls can then be placed further apart until the energy no longer changes (to a desired accuracy); this corresponds to the energy of the original system, namely the anharmonic oscillator without walls. We derive a new recursion relation to extract the coefficients up to large orders that enter the energy series expansion. The coefficients depend on a positive parameter $h>0$ that encode how far apart the walls are (where $h\to 0$ is equivalent to $L\to \infty$). With these coefficients we then calculate the ground state energy of the (quartic and sextic) anharmonic oscillator at weak, intermediate and strong coupling using the convergent series. The exact value of the ground state energy at a given coupling can be obtained directly by solving Schr\"odinger's equation numerically. At strong coupling, the result from the convergent series agrees with the exact energy to within $0.1\%$. This is an excellent match and provides strong confirmation that the perturbative approach based on finite integral limits is well suited to the strong coupling regime of a dynamical quantum system. At weak and intermediate coupling the error is completely negligible: the series matches the exact energy to within less than $0.001\%$.    

A common method that has been used for dealing with perturbative series that are asymptotic is Borel resummation (when the series is Borel summable). This has been used to study the perturbative series for the anharmonic oscillator \cite{Graffi} (see also \cite{Dillinger, Caliceti}). Borel resummation together with Pad\'e approximants have also been used to obtain the ground state energy of the anharmonic oscillator (see \cite{Marino} for details). Pad\'e approximants have recently been used in the interesting context of thermal gauge theory to interpolate between weak and strong coupling expansions \cite{Muller}. However, not all asymptotic series are Borel summable. In QED and QCD one encounters renormalons \cite{Beneke,Ioffe} where Feynman diagrams give a factorial contribution of $\approx k!$ to observables at large order $k$. Renormalons will lead in some cases to asymptotic series that are not Borel summable (e.g. the perturbative series for the anomalous magnetic moment of the electron at large order in QED \cite{Ioffe}). As already mentioned, we will encounter a simple non-Borel summable case in the next section and show that finite integral limits works (i.e. yields a convergent series in that case). This is a nice result but it is an open question if this applies to more general cases. 

Convergent expansions for the anharmonic oscillator have also been obtained by a novel rearrangement of the Hamiltonian where the unperturbed Hamiltonian is non-standard \cite{Halliday}. In \cite{Duncan} a convergent expansion for the anharmonic oscillator was obtained by extending the so-called optimized $\delta$ expansion to the finite-temperature partition function of the anharmonic oscillator.    

Our paper is organized as follows. In section 2 we consider $\lambda\,\phi^4$ theory in $0+0$ dimensions, basic integrals where the integrand contain a quadratic and quartic term with coupling $\lambda$ in the exponential. With finite integral limits, we show the perturbative series in powers of the coupling is convergent and matches the exact analytical result even when the original asymptotic series is not Borel summable. In section 3 we consider the usual asymptotic series for the ground state energy of the anharmonic oscillator and provide plots and tables of $\%$ error versus order at weak, intermediate and strong coupling where the $\%$ error is the percentage difference between the energy at order $n$ and the exact value of the energy for that coupling. In section 4, we obtain a convergent series for the ground state energy of the quartic anharmonic oscillator by placing infinite walls on both sides of the potential. In particular we derive a new recursion relation adapted to the potential with walls to extract the coefficients up to large order that enter the series. We provide plots and tables of $\%$ error versus order $n$ at weak, intermediate and strong coupling and compare the results to the asymptotic series of section 3. For all three couplings the $\%$ errors are tiny and remain that way up to large orders. In section 5 we consider the sextic anharmonic oscillator. The usual perturbative series (section 5.1) diverges even more strongly than in the quartic case. We then obtain a convergent perturbative series expansion for the energy at three different couplings (section 5.2) including strong coupling where the $\%$ error in the energy is tiny. The last section is the conclusion where we highlight our quantitative results and discuss various points of interest to potential future work.

\section{$\lambda\,\phi^4$ theory in $0+0$ dimensions: basic integral}

Consider the one-dimensional integral with quadratic plus quartic terms
\begin{align}
I=\int_{-\infty}^{\infty} dx \,e^{-\frac{1}{2} a \,x^2 -\lambda \,x^4}
\label{I}
\end{align}
where $\lambda$ is a positive real constant which will be referred to as the coupling constant. The constant $a$ is also taken to be real. It can be positive or negative and for both cases the integral is finite and yields an exact analytical expression in terms of Bessel functions. We will study the series expansions in powers of the coupling of the two cases separately. Already at this basic level, we will see that the two cases are significantly different. When dynamics are later included, we will see that the two cases correspond to qualitatively different physical systems.        

\subsection{Absolutely convergent series in powers of the coupling via finite integral limits} 

We consider below the two cases $a>0$ and $a<0$ and distinguish them with $``+"$ and $``-"$ superscripts respectively. 

\subsubsection{Case I: $a>0$}

When $a>0$, the integral $I$ given by $\reff{I}$ will be labeled $I^{+}$ and yields the exact analytical result 
\begin{align}
I^{+}=e^{\,\frac{a^2}{32 \,\lambda }} \,\sqrt{\cramped{\tfrac{a}{8\, \lambda }}} \,\,K_{\frac{1}{4}} \big(\tfrac{a^2}{32 \,\lambda }\big)\,
\label{I+}
\end{align}
where $K_n(z)$ is the modified Bessel function of the second kind.
If we replace the quartic part 
$e^{-\lambda \,x^4}$ in \reff{I} by its series expansion we obtain 
\begin{align}  
I_1^{+}&=\int_{-\infty}^{\infty} dx \,e^{-\frac{1}{2} a \,x^2}\sum_{n=0}^{\infty}
\frac{(-\lambda)^n}{n!} \,\,x^{4\,n}=\sum_{n=0}^{\infty} \frac{(-\lambda)^n}{n!} \int_{-\infty}^{\infty} dx \,e^{-\frac{1}{2} a \,x^2}\,x^{4\,n}\nonumber\\
&=\sum_{n=0}^{\infty} (-1)^n\,\frac{\lambda^n}{n!} \,\Big(\frac{2}{a}\Big)^{2n+\frac{1}{2}}\, \Gamma(2n+\tfrac{1}{2})
\label{Series1+}
\end{align}
where $\Gamma(z)$ is the gamma function and we used the result
\beq
\int_{-\infty}^{\infty} dx \,e^{-\frac{1}{2} a \,x^2}\,x^{4\,n}=\Big(\frac{2}{a}\Big)^{2n+\frac{1}{2}}\, \Gamma(2n+\tfrac{1}{2})\,.
\eeq{Gamma1}
$I_1^{+}$ is a series in powers of the coupling $\lambda$ (the $+$ superscipt denotes the case $a>0$). In switching the order of the integral and sum in the second line, each term in the series stems from an integral with infinite limits. The same result can be obtained using a generating functional which we illustrate here to make contact with QFT methods. We define the generating functional $Z[J]^{+}$ as the following one-dimensional Gaussian integral:
\beq 
Z[J]^{+}=\int_{-\infty}^{\infty} dx \,e^{-\frac{1}{2} a \,x^2 + J\,x}= \sqrt{\frac{2\,\pi}{a}}\,e^{\frac{J^2}{2\,a}}
\eeq{Gauss}
where $J$ is a real constant. We can now obtain the series $I_1^{+}$ by taking functional derivatives of $Z[J]^{+}$ with respect to $J$ and then setting $J$ to zero:  
\beq
I_1^{+}= \sum_{n=0}^{\infty} \frac{(-\lambda)^n}{n!}\Big(\frac{\delta}{\delta J}\Big)^{4n} Z[J]^{+}\Big\rvert_{J=0}=\sum_{n=0}^{\infty} (-1)^n\,\frac{\lambda^n}{n!} \,\Big(\frac{2}{a}\Big)^{2n+\frac{1}{2}}\, \Gamma(2n+\tfrac{1}{2})\,
\eeq{I1+}
which is equivalent to the series \reff{Series1+} (hence the same label $I_1^{+}$ for both).

We now consider the series $I_2^{+}$ which is obtained by replacing the infinite integration limits in the first line of \reff{Series1+} by a finite positive real number $\beta$:
\begin{align}  
I_2^{+}(\beta)&=\int_{-\beta}^{\beta} dx \,e^{-\frac{1}{2} a \,x^2}\sum_{n=0}^{\infty}
\frac{(-\lambda)^n}{n!} \,x^{4\,n}=\sum_{n=0}^{\infty} \frac{(-\lambda)^n}{n!} \int_{-\beta}^{\beta} dx \,e^{-\frac{1}{2} a \,x^2}\,x^{4\,n}\nonumber\\
&=\sum_{n=0}^{\infty} (-1)^n\,\frac{\lambda^n}{n!} \,\Big(\frac{2}{a}\Big)^{2n+\frac{1}{2}}\, \gamma(2n+\tfrac{1}{2},\tfrac{a \,\beta^2}{2})
\label{Series2+}
\end{align}
where we used the result
\beq 
\int_{-\beta}^{\beta} dx \,e^{-\frac{1}{2} a \,x^2}\,x^{4\,n}=\Big(\frac{2}{a}\Big)^{2n+\frac{1}{2}}\, \gamma(2n+\tfrac{1}{2},\tfrac{a \,\beta^2}{2})\,.
\eeq{IncGamma}
Here $\gamma(s,x)$ is the incomplete gamma function defined as
\beq
\gamma(s,x)=\int_0^x t^{s-1}\,e^{-t}\,dt \,.
\eeq{gamma}
Again, in the second line of \reff{Series2+} the order of the integral and sum were switched so that each term in the series stems from an integral but this time with finite integral limits $\beta$ in contrast to the infinite limits used to obtain series $I_1^{+}$. We can also use a generating functional to obtain $I_2^{+}(\beta)$. We define $Z[J,\beta]^{+}$ as a one-dimensional Gaussian integral with finite integral limits:
\beq 
Z[J,\beta]^{+}=\int_{-\beta}^{\beta} dx \,e^{-\frac{1}{2} a \,x^2 + J\,x}= \sqrt{\frac{2\,\pi}{a}}\,e^{\frac{J^2}{2\,a}}\, \frac{1}{2}\,\Big[\text{erf}\,\Big(\frac{J+ a \beta}{\sqrt{2\,a}}\Big)-\text{erf}\,\Big(\frac{J-a \beta}{\sqrt{2\,a}}\Big)\Big]
\eeq{GaussBeta}
where $\text{erf}\,(z)$ is the error function defined as 
\beq 
\text{erf}\,(z)=\frac{2}{\sqrt{\pi}}\int_0^z e^{-t^2}\,dt\,.
\eeq{Error}
We can therefore obtain the series $I_2^{+}(\beta)$ by taking functional derivatives of $Z[J,\beta]^{+}$ with respect to $J$ and then setting $J$ to zero:  
\beq
I_2^{+}(\beta)= \sum_{n=0}^{\infty} \frac{(-\lambda)^n}{n!}\Big(\frac{\delta}{\delta J}\Big)^{4n} Z[J,\beta]^{+}\Big\rvert_{J=0}=\sum_{n=0}^{\infty} (-1)^n\,\frac{\lambda^n}{n!} \,\Big(\frac{2}{a}\Big)^{2n+\frac{1}{2}}\, \gamma(2n+\tfrac{1}{2},\tfrac{a \,\beta^2}{2})\,.
\eeq{I2+}
The series $I_2^{+}(\beta)$, like $I_1^{+}$, is a series in powers of the coupling $\lambda$. However, it has a remarkable property: it is an absolutely convergent series for any arbitrarily large but finite $\beta$. In contrast, the series $I_1^{+}$ given by \reff{I1+} is an asymptotic series. It is simple to show that $I_1^{+}$ diverges. If the $n$th term $a_n$ in a series does not tend to zero in the limit as $n\to \infty$ then according to the divergence test, the series diverges. Since the series $I_1^{+}$ is an alternating series, we can test this by taking the limit of $|a_n|$ which yields
\beq
\lim_{n\to \infty}\,\frac{\lambda^n}{n!} \,\Big(\frac{2}{a}\Big)^{2n+\frac{1}{2}}\, \Gamma(2n+\tfrac{1}{2})\,\to \infty\,.
\eeq{DivI1}
Since the limit yields infinity instead of zero, the series $I_1^{+}$ is clearly divergent. By the dominated convergence theoerem \cite{Serone} this divergence implies that switching the order of the sum and the integral in the first line of \reff{Series1+} was not justified. The series is therefore not an exact representation of the original integral $I$ given by \reff{I} and we do not recover the analytical expression \reff{I+} (though for small $\lambda$ the series can plateau to the correct analytical value to a high accuracy before ultimately diverging \cite{Edery}).  Why is the series $I_1^{+}$ divergent when the original integral is finite? In other words, why does the switching of sum and integral in this case lead to a divergence? Note that had we expanded the quadratic part $e^{-\frac{1}{2} a \,x^2}$ in a power series and left the quartic part alone, switching sum and integral would yield an absolutely convergent series in inverse powers of $\lambda$ that yields exactly the analytical result \reff{I+}\cite{Edery}. The divergence of $I_1^{+}$ is therefore not due entirely to the switching of sum and integral. Something else is at play here. We explain this below.     

First, note that the integrand $e^{-\frac{1}{2} a \,x^2 -\lambda \,x^4}$ of the original integral $I$ is governed by the quartic part $e^{-\lambda \,x^4}$ asymptotically (as $x\to \infty$) since $x^4$ dominates over $x^2$ at large $x$. Secondly, in switching the original order of the sum and integral in the series \reff{Series1+}, we integrated to infinity before we summed to infinity. Integrating to infinity is what causes the problem. We know that $\lim\limits_{x\to\infty}\,e^{-\lambda \, x^4}$ is zero. In contrast, the series expansion of $e^{-\lambda \,x^4}$ up to any finite but arbitrary large order $N$ diverges in the asymptotic limit (i.e. $\lim\limits_{x\to\infty}\sum_{n=0}^N (-\lambda)^n\,x^{4n}/n!\to \pm\,\infty$). \textit{ Therefore the series $I_1^{+}$ cannot capture the asymptotic behavior of the original integral $I$, which is governed by $e^{-\lambda \,x^4}$}. However, if we integrate to a finite value $\beta$ there is no issue because for $x=\beta$, the series expansion of $e^{-\lambda\, x^4}$ converges to any desired accuracy if we sum enough terms. This is why the series $I_2^{+}(\beta)$ is an absolutely convergent series for any finite but arbitrarily large $\beta$. It is easy to show its absolute convergence. The $n$th term in the series \reff{I2+} is $a_n=\frac{(- \lambda)^n}{n!} \,\Big(\frac{2}{a}\Big)^{2n+\frac{1}{2}}\, \gamma(2n+\tfrac{1}{2},a \,\beta^2)$. We therefore obtain
\beq
\lim_{n\to \infty} \left|\frac{a_{n+1}}{a_n}\right|=0\,.
\eeq{RatioTest}
Since the limit is less than one, by the ratio test the series is absolutely convergent. Again, by the dominated convergence theorem \cite{Serone} we are guaranteed that the switching of sum and integral in \reff{Series2+} was justified.   Moreover, we can show that by summing the entire series $I_2^{+}(\beta)$ and then taking the infinite $\beta$ limit, we recover the analytical expression \reff{I+}. The series \reff{I2+}involves a sum over $n$ of various terms including the incomplete gamma function. The sum is convergent, but as it stands, we do not know of any analytical expression for it. However, this problem can be circumvented. The incomplete gamma function has a series expansion \cite{Gradshteyn} given by 
\beq
\gamma(s,x) =\sum_{\ell=0}^{\infty} (-1)^{\ell} \frac{x^{s+\ell}}{\ell!\,(s+\ell)}
\eeq{Seriesgamma}
so that 
\beq 
\gamma(2n+\tfrac{1}{2},\tfrac{a \,\beta^2}{2})=\sum_{\ell=0}^{\infty} (-1)^{\ell} \frac{(a \beta^2)^{2n+\tfrac{1}{2} +\ell}}{\ell!\,(2n+\tfrac{1}{2}+\ell)\,2^{2n+\tfrac{1}{2} +\ell}}\,.
\eeq{Seriesgamma2}
Substituting \reff{Seriesgamma2} into the series \reff{I2+} yields
\beq
I_2^{+}(\beta)= \sum_{\ell=0}^{\infty} \frac{(-a\,\beta^2)^{\ell}}{2^\ell \,\ell!}\sum_{n=0}^{\infty} \frac{(-\lambda)^n}{n!}\,\frac{(\beta^2)^{2n+\tfrac{1}{2}}}{2n+\tfrac{1}{2}+\ell}\,.
\eeq{I2A+}
The sum over $n$ can now be performed and yields the following analytical expression:
\beq
\sum_{n=0}^{\infty} \frac{(-\lambda)^n}{n!}\,\frac{(\beta^2)^{2\,n+\tfrac{1}{2}}}{2\,n+\tfrac{1}{2}+\ell}=\frac{1}{2}\,\gamma\,(\tfrac{\ell}{2}+\tfrac{1}{4}, \beta^4 \,\lambda)\,\lambda^{-\frac{\ell}{2}-\frac{1}{4}}\, \beta^{-2 \ell} \,.
\eeq{nsum}
Substituting \reff{nsum} into \reff{I2A+} we obtain the following sum
\beq
I_2^{+}(\beta)=\sum_{\ell=0}^{\infty} \frac{(-a)^{\ell}}{2^{\ell+1}\,\ell!}\,\gamma\,(\tfrac{\ell}{2}+\tfrac{1}{4}, \beta^4 \,\lambda)\,\lambda^{-\frac{\ell}{2}-\frac{1}{4}}\,.
\eeq{I2B}
Having completed the sum over $n$ in \reff{nsum}, we are now free to take the infinite 
$\beta$ limit before we sum over $\ell$. Only the incomplete gamma function $\gamma\,(\tfrac{\ell}{2}+\tfrac{1}{4}, \beta^4 \,\lambda)$ depends on $\beta$ and in the infinite $\beta$ limit it simply becomes the usual gamma function $\Gamma(\tfrac{\ell}{2}+\tfrac{1}{4})$. We therefore obtain  
\beq
\lim_{\beta\to \infty} I_2^{+}(\beta)=\sum_{\ell=0}^{\infty} \frac{(-a)^{\ell}}{2^{\ell+1}\,\ell!}\,\Gamma\,(\tfrac{\ell}{2}+\tfrac{1}{4})\,\lambda^{-\frac{\ell}{2}-\frac{1}{4}}=e^{\frac{a^2}{32\lambda}} \,\sqrt{\cramped{\frac{a}{8\,\lambda }}} \,\,K_{\frac{1}{4}} \Big(\frac{a^2}{32 \,\lambda }\Big)
\eeq{I2C}
where $K_n(z)$ is the modified Bessel function of the second kind. The above result is exactly the same analytical expression as \reff{I+}. We have therefore shown that $I_2^{+}(\beta)$ based on finite integral limits is an absolutely convergent series and is equal to the original integral \reff{I} in the infinite $\beta$ limit. 

We did not have to go far to encounter an asymptotic series: they arise already in $\lambda\,\phi^4$ theory in $0+0$ dimensions i.e. in the series expansion of a one-dimensional integral. Though the original integral was finite, the series $I_1^{+}$ given by \reff{I1+} diverged. Perturbative series expansions in QFT are very useful at weak coupling and are at the heart of most Standard Model calculations using Feynman diagrams. Nonetheless, they remain asymptotic series which limit their use at strong coupling. It is therefore worthwhile to summarize below in point form what led $I_1^{+}$ to be an asymptotic series.    
\begin{itemize}
\item
The integrand of the original integral has a quadratic part and a quartic part and the integral limits are infinite. Asymptotically, as $x\to\infty$, the quartic part dominates over the quadratic part. So asymptotically, the integrand of the original integral is governed by the term $e^{-\lambda \,x^4}$.

\item
The series $I_1^{+}$ is based on expanding the quartic part $e^{-\lambda \,x^4}$ which is the same part \textit{that governs the asymptotics of the original integral}. 

\item
Each term in the series $I_1^{+}$ is obtained by integrating to infinity where the series expansion of $e^{-\lambda \,x^4}$ is not valid. This means that the series $I_1^{+}$ \textit{never captures the asymptotic behaviour of the original integral} and as a consequence diverges. We saw this mathematically: the $n$th term in the series is proportional to $\int_{-\infty}^{\infty} e^{-x^2}\,\tfrac{x^{4\,n}}{n!}\, 
dx=\tfrac{\Gamma(2n+1/2)}{n!}$ which diverges as $n \to \infty$. 
\end{itemize}

We saw that the remedy to this situation is to have finite integral limits. This led to $I_2^{+}(\beta)$, which is an absolutely convergent series in powers of the coupling for any finite $\beta$. We also showed that it was equal to the analytical expression for the original integral in the infinite $\beta$ limit (with the limit taken after the series has been summed). As we will now see, the divergence of $I_1$ becomes even worse when $a<0$ which highlights the importance of the finite integral limit method used here. 

\subsubsection{Case II: $a<0$}

When $a<0$, the integral $I$ given by $\reff{I}$ is labeled $I^{-}$ and yields the exact analytical result 
\begin{align}
I^{-}=\frac{\pi}{4} \, e^{\frac{a^2}{32\, \lambda }} \sqrt{\frac{|a|}{\lambda }} \left(I_{\frac{1}{4}}\left(\frac{a^2}{32 \lambda }\right)+I_{-\frac{1}{4}}\left(\frac{a^2}{32 \lambda }\right)\right)
\label{I-}
\end{align}
where $I_n(z)$ is the modified Bessel function of the first kind. If we now replace the quartic part $e^{-\lambda \,x^4}$ in \reff{I} by its series expansion one obtains  
\begin{align}  
I_1^{-}&=\int_{-\infty}^{\infty} dx \,e^{-\frac{1}{2} a \,x^2}\sum_{n=0}^{\infty}
\frac{(-\lambda)^n}{n!} \,\,x^{4\,n}=\sum_{n=0}^{\infty} \frac{(-\lambda)^n}{n!} \int_{-\infty}^{\infty} dx \,e^{-\frac{1}{2} a \,x^2}\,x^{4\,n}\nonumber\\
&= \sum_{n=0}^{\infty} b_n\, \lambda^n
\label{Series1-}
\end{align}
where 
\beq
|b_n|= \frac{1}{n!}\int_{-\infty}^{\infty} dx \,e^{-\frac{1}{2} a \,x^2}\,x^{4\,n}\to \infty \quad \forall \,n\in \mathbb{N}_0 \, \text{and } a<0\,.
\eeq{an}
Not only is the sum of the series $I_1^{-}$ divergent, the coefficients $b_n$ for each term in the series are divergent;  this implies that it is not even Borel summable\footnote{Recall that one of the criteria for a series to be Borel summable is that $|b_n|\leq C^{n+1} \,n!$ where $C$ is a constant \cite{Strocchi,Marino}.}. This is striking since the original integral $I$ given by \reff{I} for the case $a<0$ is finite and yields the analytical expression $I^{-}$ given by \reff{I-}. In contrast to $I_1^{+}$, we cannot derive $I_1^{-}$ from a generating functional since $Z[J]^{-}=\int_{-\infty}^{\infty} dx \,e^{-\frac{1}{2} a \,x^2 + J\,x} \to \infty$ for $a<0$. Fortunately, it turns out that the same remedy employed for the asymptotic series $I_1^{+}$, namely replacing infinite integral limits by finite ones, can be applied to the more intractable case of $I_1^{-}$.     

Consider the series $I_2^{-}$ which is obtained by replacing the infinite integration limits in the first line of \reff{Series1-} by a finite positive real number $\beta$:
\begin{align}  
I_2^{-}(\beta)&=\int_{-\beta}^{\beta} dx \,e^{-\frac{1}{2} a \,x^2}\sum_{n=0}^{\infty}
\frac{(-\lambda)^n}{n!} \,x^{4\,n}=\sum_{n=0}^{\infty} \frac{(-\lambda)^n}{n!} \int_{-\beta}^{\beta} dx \,e^{-\frac{1}{2} a \,x^2}\,x^{4\,n}\nonumber\\
&=\sum_{n=0}^{\infty} (-1)^{n+1}\,\frac{\lambda^n}{n!} \,\Big(\frac{2}{|a|}\Big)^{2n+\frac{1}{2}}\, i\,\gamma(2n+\tfrac{1}{2},\tfrac{a \,\beta^2}{2})
\label{Series2-}
\end{align}
where the incomplete gamma function $\gamma(2n+\tfrac{1}{2},\tfrac{a \,\beta^2}{2})$ defined by \reff{gamma} is now purely imaginary since $\tfrac{a \,\beta^2}{2}$ is negative (the result \reff{Series2-} is real as expected since $i\,\gamma(2n+\tfrac{1}{2},\tfrac{a \,\beta^2}{2})$ is real). The series $I_2^{-}$ is a series in powers of the coupling. It can easily be shown to be an absolutely convergent series for any finite but arbitrarily large $\beta$. The $n$th term in the series is $a_n=(-1)^{n+1}\,\frac{\lambda^n}{n!} \,\Big(\frac{2}{|a|}\Big)^{2n+\frac{1}{2}}\, i\,\gamma(2n+\tfrac{1}{2},\tfrac{a \,\beta^2}{2})$ and we obtain $\lim_{n\to \infty} \left|\tfrac{a_{n+1}}{a_n}\right|=0$. Since the limit is less than one, by the ratio test, the series is absolutely convergent. 

We can define a generating functional for the case $a<0$ only by having finite integral limits:
\beq
Z[J,\beta]^{-}=\int_{-\beta}^{\beta} dx \,e^{-\frac{1}{2} a \,x^2 + J\,x}= e^{\frac{J^2}{2 a}} \,\sqrt{\frac{\pi }{-2a}}\,\left[\text{erfi}\left(\frac{J-a \beta }{\sqrt{-2 a}}\right)-\text{erfi}\left(\frac{J+a \beta }{\sqrt{-2 a}}\right)\right]\,.
\eeq{ZJBM}
In \reff{ZJBM}, $\text{erfi}\,(z)=\text{erf}\,(i\,z)/i$ where $\text{erf}\,(z)$ is the error function given by \reff{Error}. The series $I_2^{-}(\beta)$ can now be obtained via functional derivatives with respect to $J$ of the generating functional (and then setting $J$ to zero):
\beq
I_2^{-}(\beta)= \sum_{n=0}^{\infty} \frac{(-\lambda)^n}{n!}\Big(\frac{\delta}{\delta J}\Big)^{4n} Z[J,\beta]^{-}\Big\rvert_{J=0}=\sum_{n=0}^{\infty} (-1)^{n+1}\,\frac{\lambda^n}{n!} \,\Big(\frac{2}{|a|}\Big)^{2n+\frac{1}{2}}\, i\,\gamma(2n+\tfrac{1}{2},\tfrac{a \,\beta^2}{2})\,
\eeq{I2-}
which is the same result as \reff{Series2-}. We can show that the limit as $\beta\to\infty$ of $I_2^{-}(\beta)$ reproduces the analytical expression \reff{I-} corresponding to the original integral \reff{I} for $a<0$. The limit must be taken after summing the series over $n$. As it stands, we do not know any analytical expression for the summation over $n$ of the series \reff{I2-}. However, as before, we can bypass this obstacle by replacing the incomplete gamma function in \reff{I2-} by its series expansion \reff{Seriesgamma2}. This yields 
\beq
I_2^{-}(\beta)= \sum_{\ell=0}^{\infty} \dfrac{|a|^{\ell}}{\ell !\,2^{\ell}}\sum_{n=0}^{\infty}\dfrac{(-1)^n \lambda^n\,(\beta^2)^{2 n +1/2+\ell}}{n!\,(2n+\tfrac{1}{2}+\ell)}\,.
\eeq{I2-gamma}
The sum over $n$ can now be performed and yields the following expression:
\beq
\sum_{n=0}^{\infty}\dfrac{(-1)^n \lambda^n\,(\beta^2)^{2 n +1/2+\ell}}{n!\,(2n+\tfrac{1}{2}+\ell)}=\dfrac{1}{2}\, \lambda ^{-\frac{\ell}{2}-\tfrac{1}{4}}\, \gamma \left(\frac{\ell}{2}+\frac{1}{4},\beta ^4 \lambda \right)\,.
\eeq{Sum_n} 
We can now take the infinite $\beta$ limit of \reff{I2-gamma} since the sum over $n$ can now be replaced by the above analytical expression. This yields  
\begin{align}
\lim_{\beta\to\infty}I_2^{-}(\beta)&= \sum_{\ell=0}^{\infty} \dfrac{|a|^{\ell}}{\ell !\,2^{\ell+1}} \,\, \lambda ^{-\frac{\ell}{2}-\frac{1}{4}}\,\lim_{\beta\to\infty} \gamma \left(\frac{\ell}{2}+\frac{1}{4},\beta ^4 \lambda \right)\nonumber\\
&=\sum_{\ell=0}^{\infty} \dfrac{|a|^{\ell}}{\ell !\,2^{\ell+1}} \,\, \lambda ^{-\frac{\ell}{2}-\frac{1}{4}}\,\Gamma \left(\frac{\ell}{2}+\frac{1}{4}\right)\nonumber\\
&=\frac{\pi}{4} \, e^{\frac{a^2}{32\, \lambda }} \sqrt{\frac{|a|}{\lambda }} \left(I_{\frac{1}{4}}\left(\frac{a^2}{32 \lambda }\right)+I_{-\frac{1}{4}}\left(\frac{a^2}{32 \lambda }\right)\right)
\label{I2-gamma2}
\end{align}
which is exactly the analytical expression \reff{I-} of the original integral \reff{I} for the case $a<0$. Above we used $\lim_{\beta\to\infty} \gamma \left(\frac{\ell}{2}+\frac{1}{4},\beta ^4 \lambda \right)=\Gamma \left(\frac{\ell}{2}+\frac{1}{4}\right)$ where $\Gamma[z]$ is the usual gamma function.

Note that if one had expanded the quadratic part instead of the quartic part, the series would be absolutely convergent without requiring finite integral limits. The reason is that $e^{-\lambda \,x^4}$ would remain as is (not expanded) so that the series would capture correctly the asymptotic behavior of the original integral. Mathematically, the $n$th term in the series would stem from an integral like $\int_{-\infty}^{\infty} e^{-\lambda \,x^4} \,\tfrac{x^{2\,n}}{n!}\,dx$. By the ratio test, the series can be shown to be absolutely convergent. This yields a series in \textit{inverse powers of the coupling} which is particularly useful at strong coupling \cite{Edery}. We will not explore this series further in this work. Our focus will be the study of the usual perturbative series based on expanding the interaction part in powers of the coupling. 

\section{$\lambda\,\phi^4$ theory in $0+1$ dimensions: quantum anaharmonic oscilator}

We would now like to apply the ideas of the previous section to a dynamical physical system. The closest thing to a basic integral with quadratic and quartic terms is the path integral of the quantum anharmonic oscillator; this is $\lambda\,\phi^4$ theory in $0+1$ dimensions. The potential for the anharmonic oscillator is given by 
\beq
V(x) =\dfrac{1}{2}\, m \,\omega^2\,x^2 + \lambda\, x^4
\eeq{Vx}
where the first term is the usual harmonic oscillator with $\omega$ the angular frequency. This is quadratic like the kinetic term and from a field theory point of view can be incorporated as part of the free theory that yields a Gaussian path integral. The second term is then an interaction term with $\lambda$ the coupling constant. One physical quantity of interest is the ground state energy $E_0$ of the quantum anharmonic oscillator. This can be solved exactly only numerically; unlike the harmonic oscillator, there is no exact analytical solution for the energies of the anharmonic oscillator. However, we can develop a series expansion for the energy in powers of the coupling $\lambda$. We will show in the next section that this yields an asymptotic series. In QFT in $3+1$ dimensions, calculations at large order using Feynman diagrams are notoriously lengthy so that we rarely get to observe the asymptotic nature of the series. The advantage we have here is that we can generate quickly the coefficients at large orders and hence show explicitly the series diverging at various couplings. The asymptotic series causes no practical issues at weak coupling $\lambda$ since the series plateaus to a particular value (to a given number of decimal places) after a few orders and remains at that value for many orders before ultimately diverging at large orders. The value of the energy then matches the exact value obtained numerically to within a given number of decimal places. However, at strong coupling, the series fails completely: it diverges right from the start (the $\%$ error between the series and the exact energy increases with order starting at zeroth order). The asymptotic nature of the series prevents one from calculating a reliable value for the energy at strong coupling. We therefore develop an absolutely convergent series that yields the correct value of the energy at strong coupling (compared to numerical results).

\subsection{Energy coefficients via thermal partition function and Schr\"odinger recursion relation: asymptotic series}

In this section we generate the coefficients in a perturbative series expansion of the ground state energy in powers of the coupling $\lambda$. To make contact with quantum field theory, we first do this using the thermal partition function which has a Euclidean path integral representation. We then use a recursion relation derived from Schr\"odinger's equation. The latter has the advantage that coefficients at large order can easily be generated which will allow us to see explicitly the asymptotic nature of the series. 

The ground state energy of the anharmonic oscillator can be extracted from its thermal partition function. The latter has a Euclidean path integral representation and like all interacting QFTs, has no exact analytical solution. We therefore expand the path integral in powers of the coupling constant $\lambda$ and this can be calculated via functional derivatives of a generating functional $Z[J]$. $Z[J]$ is in fact the thermal partition function of the forced harmonic oscillator which has an exact analytical expression. The ground state energy can then be expanded in powers of the coupling $\lambda$ and we extract the first few coefficients (up to third order). The energy expansion can also be organized in terms of connected Feynman diagrams (see 
\cite{Marino} for a clear presentation) and the coefficients in the two cases match as they should.  
       
The thermal partition function is defined as 
\beq
Z(\beta)= \text{tr}\,e^{-\beta\,H}
\eeq{Z1}
where $H$ is the Hamiltonian and $\beta$ is related to the inverse temperature. If the system has a nondegenerate set of discrete energies $E_n$ ($n=0,1,2,...$) then the thermal partition function can be expressed as 
\beq
Z(\beta)=\sum_{n=0}^{\infty} e^{-\beta\,E_n}\,.
\eeq{Z2}
If the energies increase with $n$ such that $E_n <E_{n+1}$, the ground state energy $E_0$ can be extracted from $Z(\beta)$ by taking the infinite $\beta$ limit of its logarithm:
\beq
E_0=\lim_{\beta \to \infty}\dfrac{-1}{\beta}\, \log Z(\beta)\,.
\eeq{E0}
The thermal partition function can be expressed as a Euclidean path integral 
\beq
Z(\beta)=\int \mathcal{D}x(\tau)\, e^{-S_E}
\eeq{Z3}
where the Euclidean action $S_E$ is given by
\begin{align}
S_E=\int_{-\beta/2}^{\beta/2} \Big(\dfrac{1}{2}\, m \,\dot{x}^2 +V(x)\Big)\,d\tau\,.
\label{SE}
\end{align}
Due to the trace in \reff{Z1}, the trajectories in \reff{Z3} have periodic boundary conditions: $x(-\beta/2)=x(\beta/2)$ where $\beta$ is the total time to complete any given path. $V(x)$ is the potential and a dot above $x=x(\tau)$ is a derivative with respect to the time $\tau$. Substituting the anharmonic potential \reff{Vx} for $V(x)$ we obtain
\beq 
Z(\beta)=\int \mathcal{D}x(\tau) \,\exp\Big[\int_{\beta/2}^{\beta/2} -\Big(\dfrac{1}{2}\, m \,\dot{x}^2 +\dfrac{1}{2}\, m \,\omega^2\,x^2 + \lambda\, x^4 \Big)\,d\tau\Big] \,.
\eeq{Z4}
The above can be expanded as a series in powers of the coupling $\lambda$:
\begin{align}
Z(\beta)&=\int \mathcal{D}x(\tau) \,\exp\Big[-\int_{\beta/2}^{\beta/2}\Big(\dfrac{1}{2}\, m \,\dot{x}^2 +\dfrac{1}{2}\, m \,\omega^2\,x^2 \Big)\,d\tau\Big]\nonumber\\&\qquad\qquad\Big(1 -\lambda \,\int_{-\beta/2}^{\beta/2} x^4(\tau)\, d\tau +\,\frac{\lambda^2}{2!}\,\int_{-\beta/2}^{\beta/2} d\tau_1 \int_{-\beta/2}^{\beta/2} x^4(\tau_1)\, x^4(\tau_2) \,d\tau_2 \,+...\Big) 
\label{ZBSeries}
\end{align}
The above series can be obtained by taking functional derivatives of the generating functional $Z_{{FH}_E}[J,\beta]$ which is the thermal partition function for the forced harmonic oscillator (hence the subscript $_{FH_E}$ where $E$ stands for Euclidean). It is given by the Euclidean path integral below (with periodic trajectories $x(-\beta/2)=x(\beta/2)$):
\beq
Z_{{FH}_E}[J,\beta]=\int \mathcal{D}x(\tau) \,\exp\Big[-\int_{\beta/2}^{\beta/2} \Big(\dfrac{1}{2}\, m \,\dot{x}^2 +\dfrac{1}{2}\, m \,\omega^2\,x^2 - J(\tau) \,x\Big)\,d\tau\Big] \,.
\eeq{ZFHEJ}
It is a Gaussian path integral which can be evaluated exactly. The Euclidean path integral for the forced harmonic oscillator for a particle moving from one end point $x_a$ at time $\tau_a$ to a second end point $x_b$ at time $\tau_b$ is denoted by the Kernel $K_{{FH}_E}(x_a,\tau_a;x_b,\tau_b)$. This has been evaluated previously and is given 
by \cite{Edery}: 
\begin{align}
K_{{FH}_E}&=\bigg(\frac{m }{2\,  \pi\,\sinh(\omega\,\mathcal{T})}\bigg)^{1/2}\nonumber\\&\exp\bigg\{\frac{-\,m\,\omega}{2 \,\,\sinh (\omega\,\mathcal{T})} \Big[(x_a^2+x_b^2) \cosh (\omega\,\mathcal{T}) - 2\, x_a \,x_b - \frac{2 x_a}{m\,\omega}\int_{\tau_a}^{\tau_b}\,J(\tau) \,\sinh(\omega\,(\tau_b-\tau))\,d\tau \nonumber\\&- \frac{2 x_b}{m\,\omega}\int_{\tau_a}^{\tau_b}\,J(\tau) \,\sinh (\omega\,(\tau-\tau_a))\,d\tau\nonumber\\&-\frac{2}{m^2 \,\omega^2}\int_{\tau_a}^{\tau_b}J(\tau)\,\sinh (\omega\,(\tau_b-\tau))  \int_{\tau_a}^{\tau}\,J(\sigma)\,\sinh (\omega\,(\sigma-\tau_a))\,d\sigma\,d\tau\,\Big]\bigg\}\,.
\label{FHE}
\end{align}
where $\mathcal{T}=\tau_b-\tau_a$. To obtain $Z_{{FH}_E}$ we need to impose the periodic boundary conditions $x_a=x_b=x$ with $\tau_a=-\beta/2$, $\tau_b=\beta/2$ and  then integrate over all $x$ (the path integral \reff{ZFHEJ} includes an integral over all possible $x$). Imposing the periodic boundary conditions yields  (setting $m=\omega=1$)
\begin{align}
K_{{FH}_E}(x,\beta)&=\bigg(\frac{1}{2\,  \pi\,\sinh \beta}\bigg)^{1/2}\,\exp\bigg\{\frac{-1}{2 \,\sinh \beta} \Big[2 \,x^2 \,(\cosh \beta - 1) \nonumber\\&- 2\,x \int_{-\beta/2}^{\beta/2}\,J(\tau) \,\sinh\,(\beta/2-\tau)\,d\tau - 2 \,x \int_{-\beta/2}^{\beta/2}\,J(\tau) \,\sinh\, (\tau+\beta/2)\,d\tau\nonumber\\&-2 \int_{-\beta/2}^{\beta/2} J(\tau)\,\sinh \,(\beta/2-\tau)  \int_{-\beta/2}^{\tau}\,J(\sigma)\,\sinh \,(\sigma+\beta/2)\,d\sigma\,d\tau\,\Big]\bigg\}\,.
\label{KFHEx}
\end{align}
Integrating over all $x$ we obtain an exact analytical expression for the generating functional \reff{ZFHEJ}:
\begin{align}
Z_{{FH}_E}[J,\beta]&= \int_{-\infty}^{\infty} K_{{FH}_E}(x,\beta) \,dx\,
=\frac{1}{2\,\sinh\,(\beta/2)}\,\exp\bigg\{\frac{1}{8\,\sinh^2\,(\beta/2)\,\sinh\beta}\nonumber\\& \quad\bigg[\Big(\int_{-\beta/2}^{\beta/2}\,J(q) \,\sinh\,(\beta/2 +q)\,dq +\int_{-\beta/2}^{\beta/2}\,J(\tau) \,\sinh\, (\beta/2-\tau)\,d\tau\Big)^2 \nonumber\\&+4 \,(\cosh \beta-1)
\int_{-\beta/2}^{\beta/2} \int_{-\beta/2}^{\tau} J(\sigma)\,J(\tau)\,\sinh \,(\beta/2+\sigma)\,\sinh \,(\beta/2-\tau) \,d\sigma\,d\tau\,\bigg]\bigg\}\,.
\label{ZJExact}
\end{align}
The series \reff{ZBSeries} can now be expressed as functional derivatives of $Z_{FH_E}$ (and setting $J=0$ afterwards):
\begin{align}
Z(\beta)&= Z_{{FH}_E}\Bigr\rvert_{J = 0}- \lambda\, \int_{-\beta/2}^{\beta/2}  d\tau \dfrac{\delta^4\,Z_{{FH}_E}}{\delta J(\tau)^4}\Bigr\rvert_{J = 0}\nonumber\\
&\qquad\qquad\qquad\qquad\qquad\qquad\qquad+ \dfrac{\lambda^2}{2}\int_{-\beta/2}^{\beta/2} d\tau_1\int_{-\beta/2}^{\beta/2} d\tau_2\dfrac{\delta^8\,Z_{{FH}_E}}{\delta J(\tau_1)^4\,\delta J(\tau_2)^4}\Bigr\rvert_{J = 0}+...\nonumber\\
&= Z_{H_E} \Big(1- b_1\,\lambda + \frac{b_2}{2!}\,\lambda^2 -\frac{b_3}{3!}\,\lambda^3 +...+ (-1)^n\frac{b_n}{n!}\,\lambda^n + ...\Big)\,.
\label{ZBSeries2}    
\end{align}
$Z_{H_E}$ is obtained by setting $J=0$ in the expression \reff{ZJExact} for $Z_{{FH}_E}$. It is given by 
\beq
Z_{H_E}= Z_{FH_E}\Bigr\rvert_{J = 0}=\dfrac{1}{2\,\sinh\,(\beta/2)}
\eeq{ZHE}
which is the well known result for the thermal partition function of the harmonic oscillator (i.e. $\sum_{n=0}^{\infty} e^{-\beta \,E_n}= \sum_{n=0}^{\infty} e^{-\beta \,(n+1/2)}=\tfrac{1}{2\,\sinh\,(\beta/2)}$). To obtain the series \reff{ZBSeries2} the only thing one has to calculate are the coefficients $b_n$ which are given by 
\begin{align}
b_n=\dfrac{1}{Z_{H_E}} \int_{-\beta/2}^{\beta/2}d\tau_1d\tau_2...d\tau_n\,\,\dfrac{\delta^{4n}Z_{{FH}_E}}{\delta J(\tau_1)^4 \delta J(\tau_2)^4...\delta J(\tau_n)^4}\Bigr\rvert_{J = 0}\,.
\label{bn}
\end{align}
The coefficients $b_n$ are expressed in terms of $\beta$ and can be evaluated using the exact analytical expression \reff{ZJExact} for $Z_{{FH}_E}$. We evaluated explicitly the first three coefficients and quote below the first two ($b_3$ is too lengthy to write down):
\begin{align}
b_1&=\frac{3}{4}\, \beta  \,\coth ^2\left(\frac{\beta }{2}\right)\nonumber\\
b_2&=\frac{3}{128} \,\beta \, \text{csch}^4\left(\frac{\beta }{2}\right) (81 \beta +80 \sinh (\beta )+28 \sinh (2 \beta )+60 \beta  \cosh (\beta )+3 \beta  \cosh (2 \beta ))\,.
\label{bees}
\end{align}
Inserting the series \reff{ZBSeries2} for the thermal partition function $Z(\beta)$ into the ground state energy \reff{E0} one obtains 
\begin{align}
E_0 &= \lim_{\beta \to \infty}\dfrac{-1}{\beta}\, \log Z(\beta)\nonumber\\
&=\lim_{\beta \to \infty}\dfrac{-1}{\beta}\log \Big[Z_{H_E} \big(1- b_1\,\lambda + \frac{b_2}{2!}\,\lambda^2 -\frac{b_3}{3!}\,\lambda^3 +...\big)\Big]\,.
\label{E0A}
\end{align}
Expanding the above logarithm in a series about $\lambda=0$, the ground state energy can be expressed as a series in powers of $\lambda$: 
\begin{align}
E_0=\frac{1}{2} +\sum_{j=1}^{\infty} a_j \, \lambda^j
\label{E02}
\end{align}
where the numerical coefficients $a_j$ can be expressed in terms of the $b_n$ coefficients. The first term $1/2$ stems from  
\begin{align}
\lim_{\beta \to \infty}\dfrac{-1}{\beta}\log Z_{H_E}=\lim_{\beta \to \infty}\dfrac{1}{\beta}\log \,[\,2 \sinh\,(\beta/2)\,]=\dfrac{1}{2}\,.
\end{align}
We have quoted $b_1$ and $b_2$ in \reff{bees} and we also evaluated $b_3$ (which we do not show). The first three numerical coefficients $a_j$ can be evaluated from them:
\begin{align}
a_1 &=\lim_{\beta \to \infty}\dfrac{b_1}{\beta} = \dfrac{3}{4} \nonumber\\
a_2&=\lim_{\beta \to \infty}\dfrac{b_1^2-b_2}{2\,\beta} =-\dfrac{21}{8}\nonumber\\
a_3&=\lim_{\beta \to \infty}\dfrac{2b_1^3-3 b_1\,b_2+b_3}{6\,\beta}=\dfrac{333}{16}\,.
\label{as}
\end{align}
The above coefficients are in agreement with calculations using connected Feynman diagrams \cite{Marino} up to order $\lambda^3$. Solving for $a_4$ and higher coefficients becomes somewhat lengthy. This would require calculating $b_4$ and higher coefficients $b_n$ using \reff{bn} and this becomes computationally intensive. If one uses connected Feynman diagrams, the number of diagrams grows factorially at large order $j$ so that the number of calculations become extremely large at higher orders. 

Luckily we can generate the coefficients $a_j$ quickly at large order $j$ via a recursion relation derived from Schr\"odinger's equation. We generate the coefficients up to order $j=50$. We use these coefficients to make plots and tables of the energy at weak and strong coupling to illustrate in a transparent fashion the nature of the asymptotic series. The recursion relation is known and could simply be stated \cite{Wu2, Marino}). However, in the next section we obtain an absolutely convergent series based on ideas that lead to a novel modified version of that recursion relation. So it will be worthwhile to give a brief derivation of the recursion relation here as this will be useful for the next section. 

Schr\"odinger's equation for the anharmonic oscillator is given by (setting $\hbar=m=\omega=1$):
\beq
\Big(-\dfrac{1}{2} \dfrac{d^2}{dx^2} +\dfrac{x^2}{2} + \lambda\, x^4\Big) \psi(x)= E_0\,\psi(x)\,.
\eeq{Schro}
We write the wavefunction $\psi(x)$ as a series expansion in powers of $\lambda$:  
\beq
\psi(x)= e^{-x^2/2}\sum_{m=0}^{\infty} \lambda^m\,B_m(x)
\eeq{Ansatz}
where $e^{-x^2/2}$ is the wavefunction (up to a normalizing factor) for the harmonic oscillator ($\lambda=0$) and $B_m(x)$ are polynomial functions with $B_0(x)=1$. Substituting the expansions \reff{Ansatz} and \reff{E02} for the wavefunction and energy respectively into \reff{Schro} we obtain the following equation:
\beq
x\,B_n^{'}(x) -\dfrac{1}{2}\, B_n^{''}(x) + x^4 \,B_{n-1}(x) =\sum_{p=0}^{n-1} a_{n-p} B_p(x)\,.
\eeq{RecurF}
We can express the functions $B_n(x)$ as polynomials of degree $4n$ containing only even powers of $x$ :
\beq
B_m(x)=\sum_{j=0}^{2 m} (-1)^m x^{2\,j}\, B_{m,j} 
\eeq{RecurB}     
where we define $B_{0,0}=1$ and $B_{k,0}=0$ for $k\ne 0$ so that $B_0(x)=1$ and 
$B_m(x)$ contains no constants for $m\ne 0$. 

In \reff{RecurF}, the first term on the right hand side at $p=0$ is equal to the constant $a_n$. The only constant on the left hand side is $(-1)^{n+1}\, B_{n,1}$ which stems from inserting the $j=1$ contribution $(-1)^n\, x^2 \,B_{n,1}$ into $-\tfrac{1}{2}\, B_n^{''}(x)$ . We therefore have 
\beq
a_n=(-1)^{n+1}\,B_{n,1}\,.
\eeq{Bn1}
So we need to solve for the coefficients $B_{n,1}$ as those are the ones that enter into the energy expansion. Substituting \reff{RecurB} and \reff{Bn1} into the equation \reff{RecurF} we obtain the following recursion relation for the coefficients $B_{i,j}$ \cite{Marino}:
\beq
2\,j B_{i,j}- (j+1) (2j+1) B_{i, j+1} - B_{i-1,j-2}= - \sum_{p=1}^{i-1} B_{i-p,1}\,B_{p,j}\,.
\eeq{Recursion}
In the above recursion $i \ge 1$ and $j \ge 1$. A coefficient $B_{m,k}$ is zero if $k>2\,m$ or $k<0$. Recall that $B_{0,0}=1$ and $B_{k,0}=0$ for $k \ne 0$. So we already know that the coefficient $B_{i-1,j-2}$ is zero for $j=1$, unity when $j=2$ and $i=1$ and zero when $j=2$ and $i>1$. We used the above recursion relation to generate $B_{n,1}$ from $n=1$ to $n=50$. In other words, we obtained the first $50$ coefficients $a_n$ that enter the energy expansion \reff{E02}. This enabled us to see what happens at large order.  

\begin{figure}[t]
	\centering
		\includegraphics[scale=0.8]{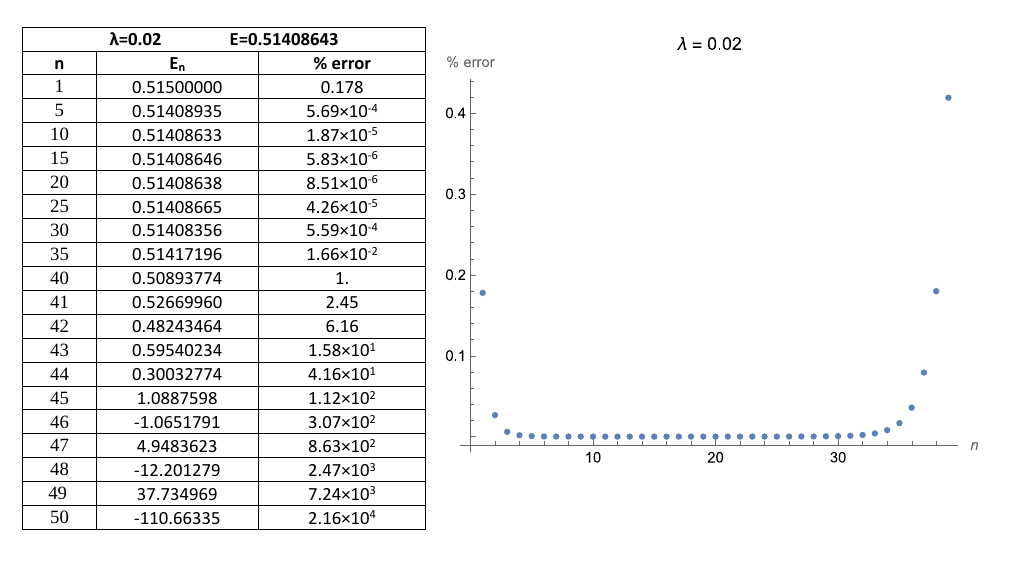}
		\caption{Results at small (weak) coupling $\lambda=0.02$. The ground state energy $E_n$ at order $n$ is quoted to eight decimal places. The table contains the $\%$ error between the energy $E_n$ and the exact energy $E=0.51408643$ obtained numerically and quoted also to eight decimal places at the top of the table. The $\%$ error is completely negligible over a long range of orders (from roughly $n=5$ to $n=35$). This is the long plateau region on the plot and implies that the series can be used to make reliable predictions. After the plateau region, the series begins to diverge and reaches a very large $\%$ error of order $10^4$ at $n=50$. We therefore have an asymptotic series but because the coupling is small, this becomes only apparent starting at very large orders (roughly starting at order $n=40$).}     
		\label{Lambda002}
\end{figure}

\begin{figure}[t]
	\centering
		\includegraphics[scale=0.8]{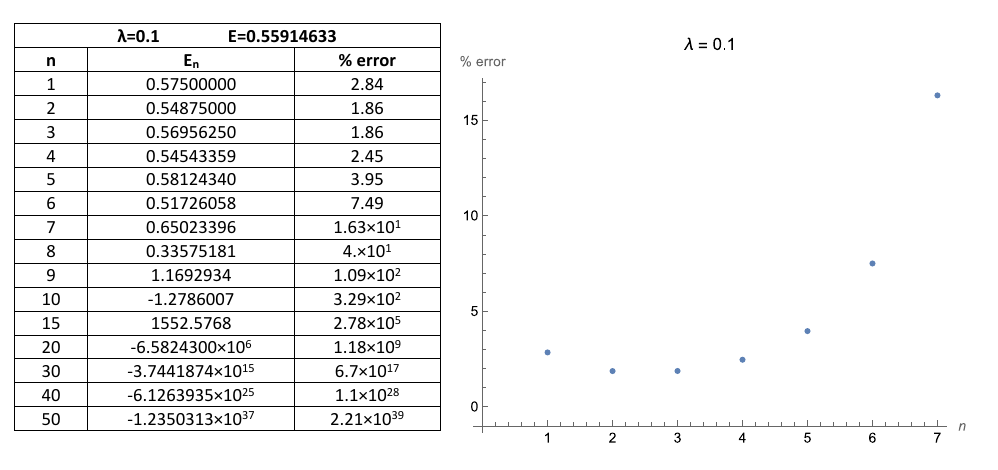}
		\caption{Results at intermediate coupling $\lambda=0.1$. The $\%$ error between the series $E_n$ and the exact value $E=0.55914633$ starts at a reasonably low value of $2.84\%$, dips to a minimum of $1.86 \%$ over the next two orders and quickly diverges afterwards (asymptotic series). So the series gets reasonably close to the exact value early on (at low orders) but in contrast to the weak coupling case, there is no plateau region where it settles close to the exact value over a long range of orders. This is why we refer to it as an ``intermediate" coupling. In the next section, we obtain an absolutely convergent series that settles/converges to the correct value with negligible $\%$  error all the way to large order $n$.}     
		\label{Lambda01}
\end{figure}

\begin{figure}[t]
	\centering
		\includegraphics[scale=0.8]{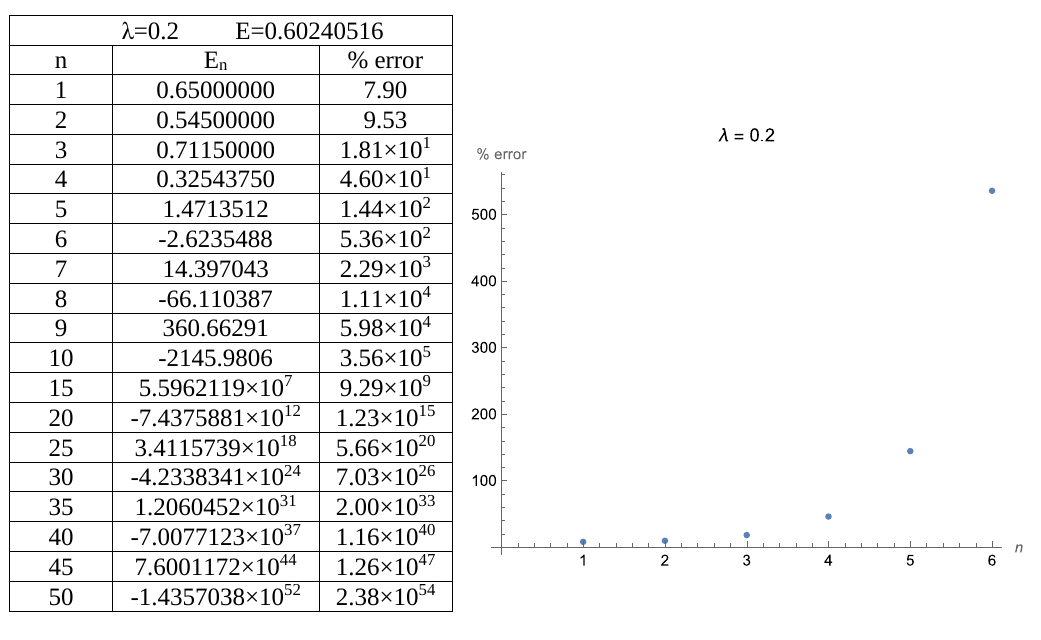}
		\caption{Results at large (strong) coupling $\lambda=0.2$. The $\%$ error starts at $8\%$ but only increases afterwards. The series never gets close to the exact value and then diverges from it very rapidly. We see that at strong coupling, the perturbative series breaks down completely and a new approach is required. It the next section 
we develop an absolutely convergent perturbative series which yields excellent results at strong coupling.}     
		\label{Lambda022B}
\end{figure}

We obtained results at three separate couplings: weak($\lambda=0.02$), intermediate ($\lambda=0.1$) and strong ($\lambda=0.2$). For each coupling, we calculated the energy $E_n$ at each order $n$ using \reff{E02} from $n=1$ to $n=50$.  We also calculate the $\%$ error between $E_n$ at a given order and the exact value $E$ for the energy obtained by solving numerically Schr\"odinger's equation. The results are presented in table form and in a plot for each coupling in figures \ref{Lambda002}, \ref{Lambda01} and \ref{Lambda022B}. The case $\lambda=0.02$ converges to the exact value over a large range of orders before ultimately diverging. The plot shows a long plateau region where it matches the exact value to within five or six decimal places (equivalent to less than $0.001 \%$ error) before it begins to diverge at larger orders (starting roughly at $n>35$). The plateau region appears because $\lambda=0.02$ is small enough (weak coupling). The advantage of being able to go to large orders, is that we are able to observe explicitly that the series is an asymptotic series. Despite this, the plateau region implies the series is still useful in the sense that one can extract from it practical results that agree with experiment (where ``experiment" in this context refers to the exact value of the energy).  At intermediate coupling $\lambda=0.1$ the series dips close to the exact value after a few orders ($\%$ error reaches a minimum of $1.86 \%$) but then quickly diverges afterwards. In contrast to the weak coupling case, there is no plateau region where it settles close to the exact value over a long range of orders. This is why we refer to it as an ``intermediate" coupling. At strong coupling $\lambda=0.2$ the $\%$ error increases with order the entire time. The series never gets close to the exact value at any order and diverges from it very rapidly. The perturbative series breaks down completely at strong coupling. For the anharmonic oscillator, the minimal error from optimal truncation (see \cite{Marino}) is expected to occur at order $n_M\approx 1/(3 \lambda)$. For $\lambda=0.02, 0.1$ and $0.2$ this occurs at $n_M \approx 17, 3$ and $2$ respectively which matches very closely the order at which the minimum error occurs in the data tables of figures \ref{Lambda002}, \ref{Lambda01} and \ref{Lambda022B} respectively. In the next section, for all three couplings, we obtain an absolutely convergent series that settles/converges to the correct value with tiny $\%$ error all the way to large order $n$.
  

\section{Convergent perturbative series and the energy of\\ the quartic anharmonic oscillator at strong coupling}

We now place infinite walls at $x=\pm L$ in the anharmonic potential where $L$ is an arbitrary finite length (we \textit{formally} recover the original potential in the limit $L\to \infty$). This means the ground state wavefunction $\psi(x)$ decreases to zero at $x\pm L$ (and remains zero for $|x|>L$). This must occur for any value of $\lambda$ including  $\lambda=0$. So we first solve the ground state wavefunction for the harmonic case which we label $\psi_0(x)$. Without walls the ground state wavefunction is $e^{-x^2/2}$ so with walls it is convenient to express it as
as
\beq
\psi_0(x) =e^{-x^2/2}\, g(x)
\eeq{HarmonicWalls}
where $g(x)$ must be zero at $x=\pm L$ and be an even function of $x$ since $\psi_0(x)$ is an even function of $x$. Substituting this into Schr\"odinger's equation we obtain 
\begin{align}
&\Big(-\dfrac{1}{2} \dfrac{d^2}{dx^2} +\dfrac{x^2}{2}\,\Big)\,\psi_0(x)= E_{\lambda=0}\,\psi_0(x)\nonumber \\
&\dfrac{1}{2}\, e^{-x^2/2}\, \Big(g(x) + 2\,x\,g'(x) - g"(x)\Big)=E_{\lambda=0}\,e^{-x^2/2}\, g(x)\,.
\label{Schro2}
\end{align}
So we obtain the differential equation 
\beq
2\,x\,g'(x)-g''(x)= h \,g(x)
\eeq{diff3}
where $h$ is a constant. The general solution is a linear combination of two functions:
\beq 
g(x)= c_1\, H_{h/2}(x) + c_2 \,\,{}_1F_1\big(-\frac{h}{4}\,;\,\frac{1}{2}\,;\, x^2\,\big)
\eeq{gg}
where $H_{h/2}(x)$ is a Hermite polynomial and ${}_1F_1\big(-\frac{h}{4},\frac{1}{2}, x^2\,\big)$ is the Kummer confluent hypergeometric function. However, the Hermite polynomial has to be excluded because it does not satisfy the required boundary condition on $g(x)$, namely that an $L$ exists where the function is zero at $x=\pm L$. In other words, for a continuous set of values of $h$, either positive or negative, the Hermite polynomial will not in general cross zero symmetrically on both sides at an equal distance (labeled $L$) to the right and left of the origin. In contrast, the hypergeometric function $\,{}_1F_1\Big(-\frac{h}{4},\frac{1}{2}, x^2\big)$ is an even function of $x$ which satisfies the boundary condition on $g(x)$ for all positive values of $h$. At $h=0$, the function is a constant equal to unity. For positive values of $h$, it is unity near the origin $x=0$ and decreases symmetrically on both sides to cross zero at values $x=\pm L$ where $L$ is determined by $h$. For example, if $h=0.001$, the function looks like figure 4. So here $L$ turns out to be close to $3$ (though the function here becomes negative for $|x|>L$ it is only defined in the range $|x|\le L$ and would be normalized within that range). Since our solution is expressed in terms of $h$, we work with $h$ instead of $L$. The case $h=0$ is equivalent to having no walls (in effect $L\to \infty$) and as $h$ increases, $L$ decreases. So the particle gets more confined as $h$ increases and we expect the ground state energy to increase with $h$.       
\begin{figure}[t]
	\centering
		\includegraphics[scale=0.8]{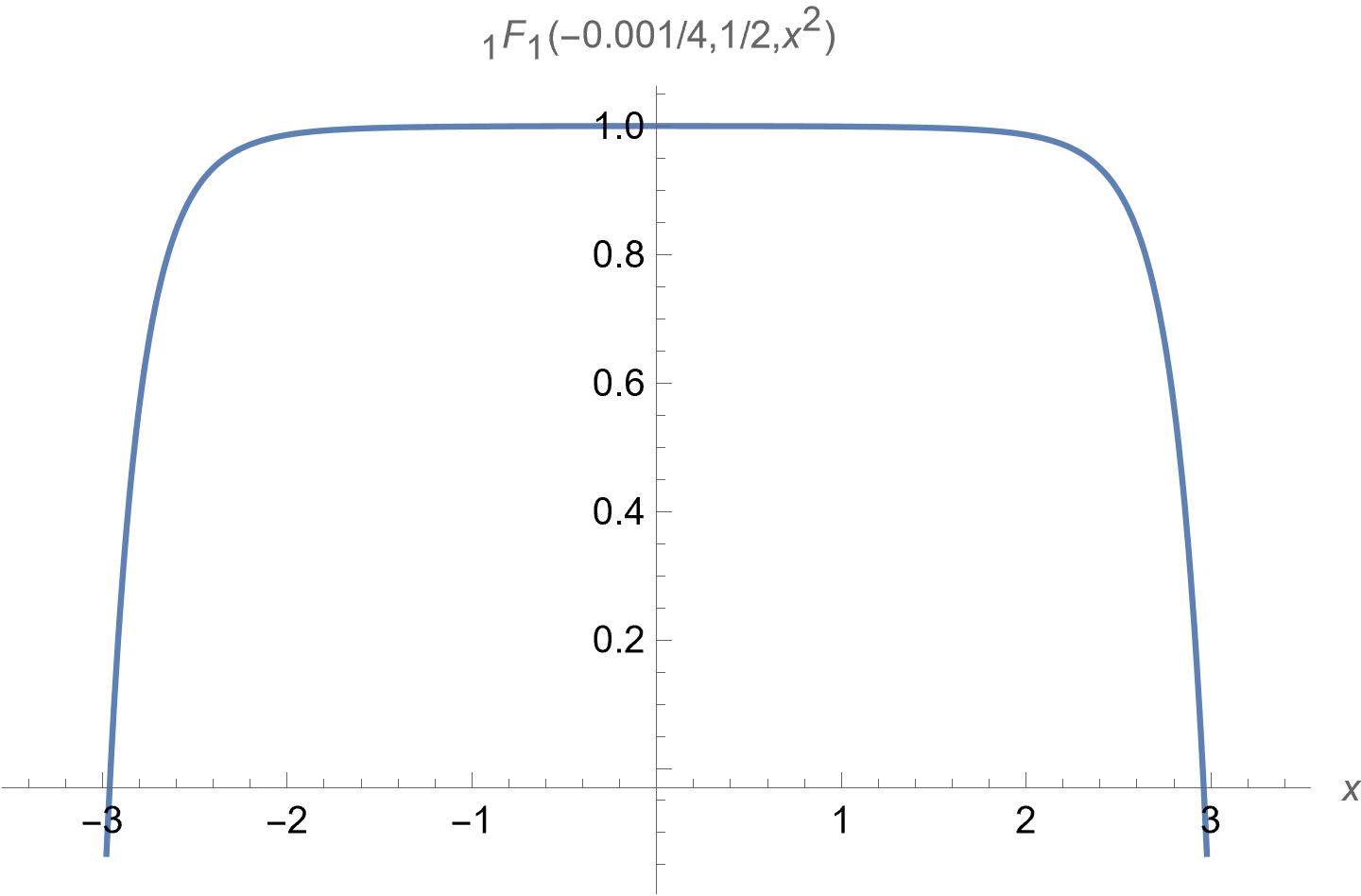}
		\caption{Plot of the Kummer confluent hypergeometric function ${}_1F_1\big(-\frac{h}{4},\frac{1}{2}, x^2\,\big)$ as a function of $x$ for $h=0.001$. The function is basically unity until it drops off on both sides to zero at $x\approx \pm 3$.}     
		\label{H1F1}
\end{figure}

The wavefunction of the harmonic oscillator with the condition that it goes to zero at $x=\pm L$ is therefore given by 
\beq
\psi_0(x) =e^{-x^2/2}\,{}_1F_1\big(-\frac{h}{4}\,;\,\frac{1}{2}\,; \,x^2\,\big)
\eeq{WallsHO}
where $h$ is positive and its value determines $L$. Substituting \reff{diff3} into \reff{Schro2} the ground state energy of the harmonic oscillator as a function of $h$ is given by
\beq
E_{\lambda=0}= \dfrac{1+h}{2}\,.
\eeq{Eh}
This is a nice simple result. We recover the energy of $1/2$ when there are no walls ($h=0$) and we see that the energy increases with $h$ as anticipated. 

Similar to \reff{Ansatz}, we express the ground state wavefunction $\psi(x)$ of the anharmonic oscillator with walls as a series expansion in powers of $\lambda$ with the zeroth order being the harmonic solution \reff{WallsHO}:
\beq
\psi(x)= e^{-x^2/2}\,{}_1F_1\big(-\frac{h}{4}\,;\,\frac{1}{2}\,;\, x^2\,\big)\,\sum_{n=0}^{\infty} \lambda^n\,B_n(x)
\eeq{Ansatz2} 
where $B_n(x)$ are again polynomials in even powers of $x$ with $B_0(x)=1$. Note that the boundary conditions at the walls are automatically satisfied i.e. $\psi(x)=0$ at $x=\pm L$ (where $L$ is determined by $h$). The ground state energy of the anharmonic oscillator with walls is now a function of $h$ ($E_0=E_0(h)$) and we can again express it as a series expansion in powers of $\lambda$:
\beq
E_0= \dfrac{1+h}{2} +\sum_{k=1}^{\infty} a_k \, g^k 
\eeq{E0h}
where we used the result \reff{Eh}. The coefficients $a_k=a_k(h)$ now depend on $h$. We recover the coefficients and energy expansion of the previous section (no walls) in the limit $h\to 0$. Our goal now is to derive a new recursion relation that will allow us to calculate the coefficients $a_k$ for a given $h$. This will be a modified version of our previous recursion relation \reff{Recursion}. Schr\"odinger's equation for the anharmonic oscillator is given by \reff{Schro}: 
\beq
\Big(-\dfrac{1}{2} \dfrac{d^2}{dx^2} +\dfrac{x^2}{2} + \lambda\, x^4\Big) \psi(x)= E_0\,\psi(x)\,.
\eeq{SchroA}
Inserting \reff{Ansatz2} and \reff{E0h} into the above, the left hand side of the equation yields
\begin{align}
\sum_{n=0}^{\infty} e^{-\frac{x^2}{2}} \lambda^n &\bigg[ (2+h) \, _1F_1\big(-\frac{h}{4}\,;\,\frac{3}{2}\,;\,x^2\,\big)\, x\,B_n'(x)\nonumber\\&+\, _1F_1\big(-\frac{h}{4}\,;\,\frac{1}{2}\,;\,x^2\,\big) \Big(\,\frac{1+h}{2}\, B_n(x) + \lambda\,  x^4 \,B_n(x)- x \,B_n'(x)-\frac{B_n''(x)}{2}\,\Big)\bigg]
\label{LHS}
\end{align}
and the right hand side yields 
\begin{align}
\sum_{n=0}^{\infty} e^{-\frac{x^2}{2}} \lambda^n \, _1F_1\big(-\frac{h}{4}\,;\,\frac{1}{2}\,;\,x^2\,\big) \Big(\,\frac{1+h}{2}\, B_n(x) + \sum_{p=0}^{n-1} a_{n-p}\,B_p(x)\,\Big)\,.
\label{RHS}
\end{align}
Note that there are two hypergeometric functions: one with $1/2$ as its second argument and the other with $3/2$. For simplicity of notation and clarity it is convenient to define the functions
\begin{align}
F_1&= \,_1F_1\big(-\frac{h}{4}\,;\,\frac{1}{2}\,;\,x^2\,\big)\nonumber \\
F_2&=\, _1F_1\big(-\frac{h}{4}\,;\,\frac{3}{2}\,;\,x^2\,\big)\,.
\label{F1F2}
\end{align}
Equating \reff{LHS} with \reff{RHS} and matching powers of $\lambda$ we obtain the following differential equation:
\begin{align}
(2+h) \,x\,B_n'(x)\,F_2 + F_1\,\big(- x \,B_n'(x)-\frac{B_n''(x)}{2}+\,x^4 \,B_{n-1}(x)\,\big)= F_1\,\sum_{p=0}^{n-1} a_{n-p}\,B_p(x) \,.
\label{Dif2}  
\end{align} 
As a simple check, when $h=0$, $F_1$ and $F_2$ are both unity and we recover our previous differential equation \reff{RecurF} as expected.
As before, we write down $B_n(x)$ as polynomials of degree $4n$ containing only even powers of $x$ :
\beq
B_n(x)=\sum_{j=0}^{2 n} (-1)^n x^{2\,j}\, B_{n,j}\,.
\eeq{BPoly}     
Again, we define $B_{0,0}=1$ and $B_{k,0}=0$ for $k\ne 0$ so that $B_0(x)=1$ and $B_m(x)$ contains no constants for $m\ne 0$. The hypergeometric functions $F_1$ and $F_2$ defined by \reff{F1F2} have the following series expansions:
\begin{align}
F_1&=\sum_{m=0}^{\infty} c_m \dfrac{(x^2)^m}{m!}=1+ c_1\, x^2 +\frac{1}{2!} \,c_2\,x^4 +...
\nonumber\\
F_2&=\sum_{m=0}^{\infty} b_m \dfrac{(x^2)^m}{m!}==1+ b_1\, x^2 +\frac{1}{2!}\, b_2\,x^4 +...
\label{Fifa}
\end{align}
where $c_m$ and $b_m$ are coefficients given by 
\begin{align}
c_m&=\dfrac{\Gamma(m-h/4)\, \Gamma(1/2)}{\Gamma(-h/4) \,\Gamma(m+1/2)}\nonumber\\
b_m&=\dfrac{\Gamma(m-h/4)\, \Gamma(3/2)}{\Gamma(-h/4) \,\Gamma(m+3/2)}\,.
\label{cb}
\end{align}
Note that $c_0=b_0=1$ regardless of the value of $h$ including the limit $h\to 0$. Also, $\lim_{h\to 0} b_m= \lim_{h\to 0} c_m =0$ for $m \ge 1$.  The first term in the series expansion of $F_1$ is unity. This implies that in \reff{Dif2}, the series expansion on the right hand side contains the constant $a_n$. The only constant in the series expansion on the left hand side stems from the term $-F_1\,\tfrac{B_n^{''}(x)}{2}$. The $j=1$ contribution to $B_n(x)$ given by \reff{BPoly} is $(-1)^n\, x^2 \,B_{n,1}$. When this is inserted into $-F_1\,\tfrac{B_n^{''}(x)}{2}$ one obtains $F_1\,(-1)^{n+1}\, B_{n,1}$ whose series expansion contains the constant $(-1)^{n+1}\, B_{n,1}$ since the first term in the expansion of $F_1$ is unity. We therefore obtain again 
\beq
a_n= (-1)^{n+1}\, B_{n,1}\,.  
\eeq{an3}
Therefore, the problem of finding the ground state energy via the series expansion \reff{E0h} is reduced to solving for the coefficients $B_{n,1}$ which will now depend on $h$. 
Substituting \reff{an3}, \reff{BPoly} and \reff{Fifa} into \reff{Dif2} we finally obtain our new recursion relation:
\pagebreak
\begin{align}
\sum_{\substack{j=1\\m=k-j\\ 1\le k \le 2n}}^{k}& \Big[2\,j B_{n,j}\Big((2+h)\,\frac{b_m}{m!}-\frac{c_m}{m!}\Big) -(j+1)\,(2j+1) B_{n,j+1} \frac{c_m}{m!}\nonumber\\
&\qquad- B_{n-1,j-2}\,\frac{c_m}{m!}+\sum_{p=1}^{n-1} B_{n-p,1}\,B_{p,j}\,\frac{c_m}{m!}\Big]=0\,.
\label{Recursionh}
\end{align}
The above recursion relation yields one equation for each $k$ where $k$ runs from $1$ to $2 n$. At each $k$, the equation is composed of a sum of terms from $j=1$ to $k$ with $m=k-j$. At each order $n$, one can then solve for the coefficient $B_{n,1}$ by solving sequentially the set of $2n$ equations. The coefficients $B_{n,1}$ are a function of $h$ and hence so are the coefficients $a_n=(-1)^{n+1}\,B_{n,1}$ which enter into the energy expansion \reff{E0h}. To illustrate we write down explicitly below the first two coefficients $a_n$ as a function of $h$:
\begin{align}
a_1&=\frac{3}{4 + 12\,h + 6\,h^2}\nonumber\\
a_2&=-\frac{9 \left(33 h^3+99 h^2+94 h+28\right)}{\left(3 h^2+6 h+2\right)^2 \left(155 h^4+620 h^3+780 h^2+320 h+24\right)}\,.
\label{a1a2h}
\end{align}
As a simple check on the above formulas, we recover our previous $a_1$ and $a_2$ coefficients given by \reff{as} if we set $h=0$ (no walls): 
\beq 
a_{1\,|_{h=0}}= \frac{3}{4} \text{ and } a_{2\,|_{h=0}}=-\frac{21}{8}\,.
\eeq{a1a2h0}
To generate quickly the $a_n$ at large order $n$ it is much faster to set a numerical value for $h$ before solving the equations. One then obtains a numerical value for the coefficients $a_n$ for a particular numerical value of $h$.    

The parameter $h$ appears in the new recursion relation \reff{Recursionh} explicitly and via the coefficients $b_m$ and $c_m$ given by \reff{cb}. We can easily verify that it reduces to our previous recursion relation \reff{Recursion} in the limit $h\to 0$ (no walls). In that limit, the coefficients $b_m$ and $c_m$ are zero except at $m=0$ where $b_0=c_0=1$. Since $m=k-j$ this means that the sum over $j$ is non-zero only at $j=k$. The sum over $j$ in \reff{Recursionh} can therefore be removed and $j$ replaced by $k$ in the equation. This yields our previous recursion relation \reff{Recursion} (our $k$ is the $j$ appearing in \reff{Recursion} and they can take on integer values between $1$ and $2\,n$ inclusively). This provides a further check on our new recursion relation and is consistent with the results presented in \reff{a1a2h0}. 

When the coefficients $a_n$ are now substituted into the energy expansion \reff{E0h}, the series is no longer an asymptotic series but an \textit{absolutely convergent} series for any finite positive value of $h$ ($h\ne 0$). For a given $h$, the series converges to a particular energy (to within a given number of decimal places) with more terms required to reach convergence if $h$ is smaller. When decreasing $h$ no longer changes the energy to a desired level of accuracy, one is then close to the exact energy (i.e. close to within that accuracy).

\begin{figure}[t]
	\centering
		\includegraphics[scale=0.8]{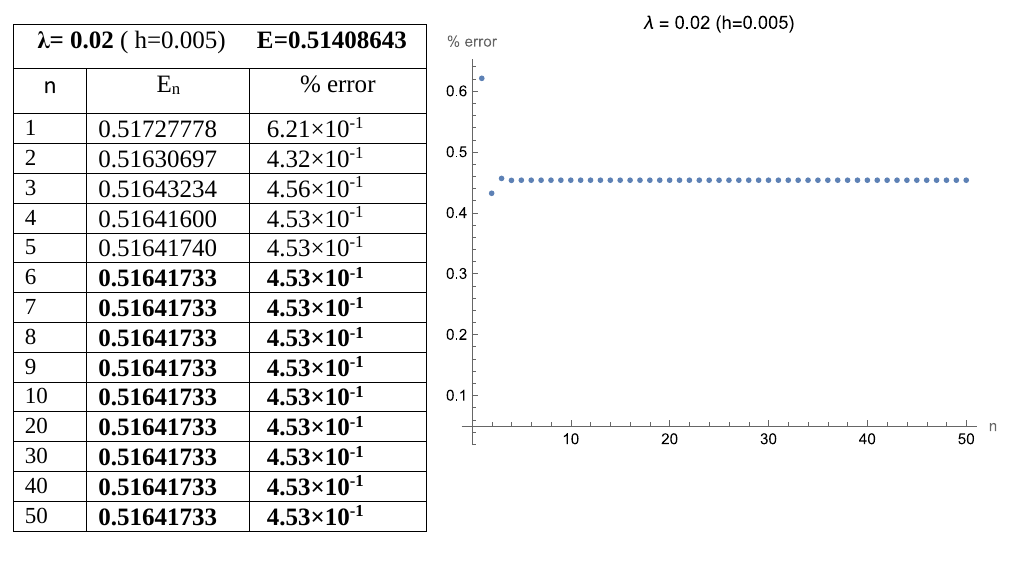}
		\caption{Results at weak coupling $\lambda=0.02$ with $h=0.005$. The series converges to the exact value of the energy to within less than $0.5\%$ up to arbitrary large orders (we show it here up to $n=50$). It is an absolutely convergent series.  In contrast, the usual perturbative series at weak coupling diverges at large order as in the plot of Fig. \ref{Lambda002}.}     
		\label{Lambda002A}
\end{figure}

\begin{figure}[t]
	\centering
		\includegraphics[scale=0.8]{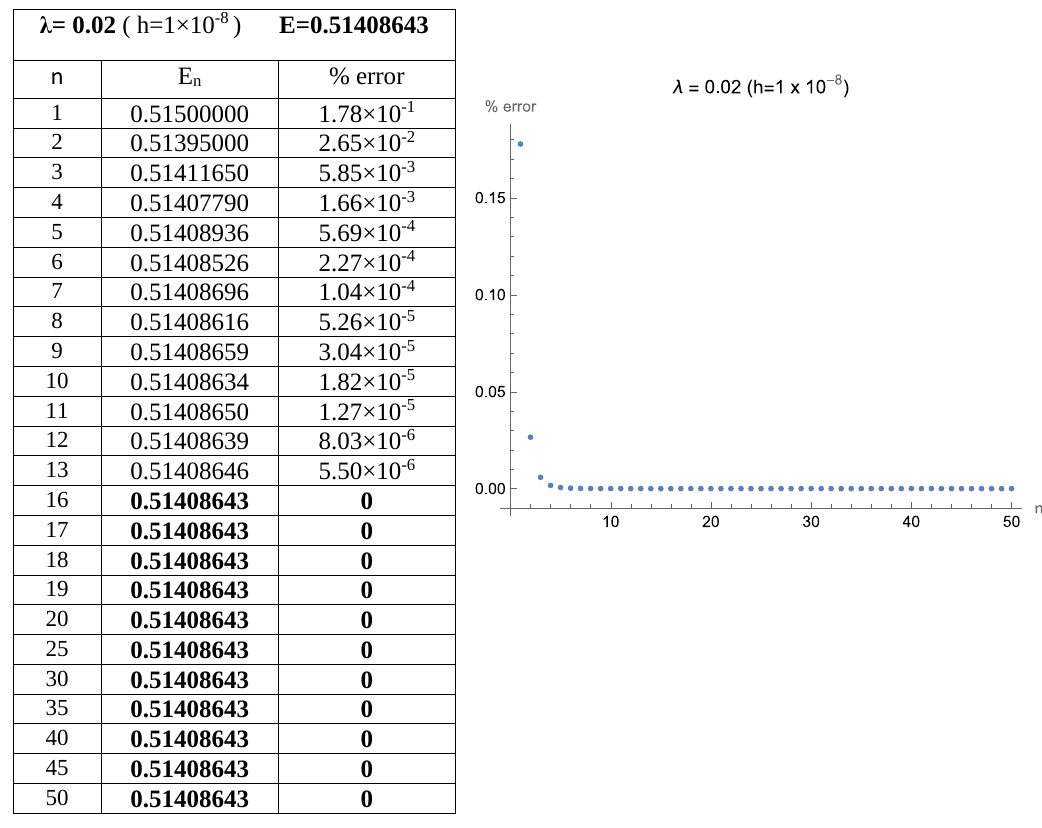}
		\caption{Results at weak coupling $\lambda=0.02$ with $h=1\times 10^{-8}$. The parameter $h$ is smaller by a factor of $5\times 10^5$ compared to the previous case with $h=0.005$. The series now matches the correct value to eight decimal accuracy (hence no error to within eight decimal places). It converges up to arbirary large orders (shown here up to $n=50$). Though the parameter $h$ is tiny, it is not zero, and this is why it yields an absolutely convergent series instead of an asymptotic one.}      
		\label{Lambda002B}
\end{figure}

\begin{figure}[t]
	\centering
		\includegraphics[scale=0.8]{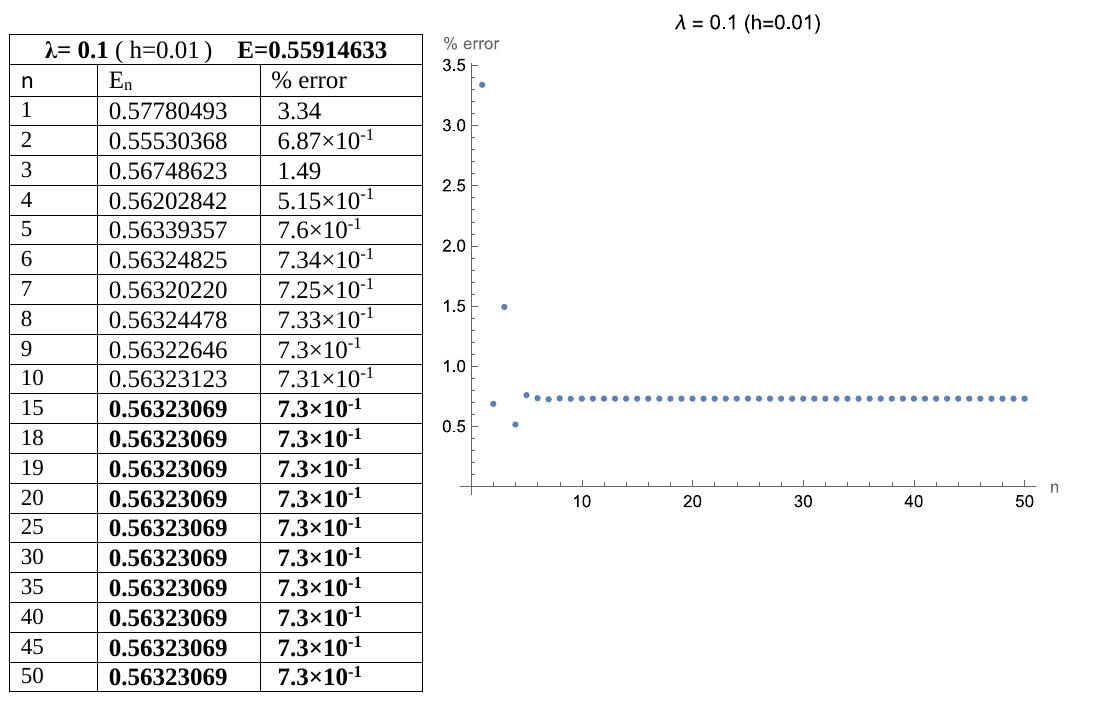}
		\caption{Results at intermediate coupling $\lambda=0.1$ with $h=0.01$. The series converges to the exact value of the energy to within less than $0.8\%$ up to arbitrary large orders (we show it here up to $n=50$). Recall that in contrast, the plot for the usual perturbative series at this coupling diverged quickly after an initial dip (see Fig.\ref{Lambda01}). }     
		\label{Lambda001A}
\end{figure}

\begin{figure}[t]
	\centering
		\includegraphics[scale=0.8]{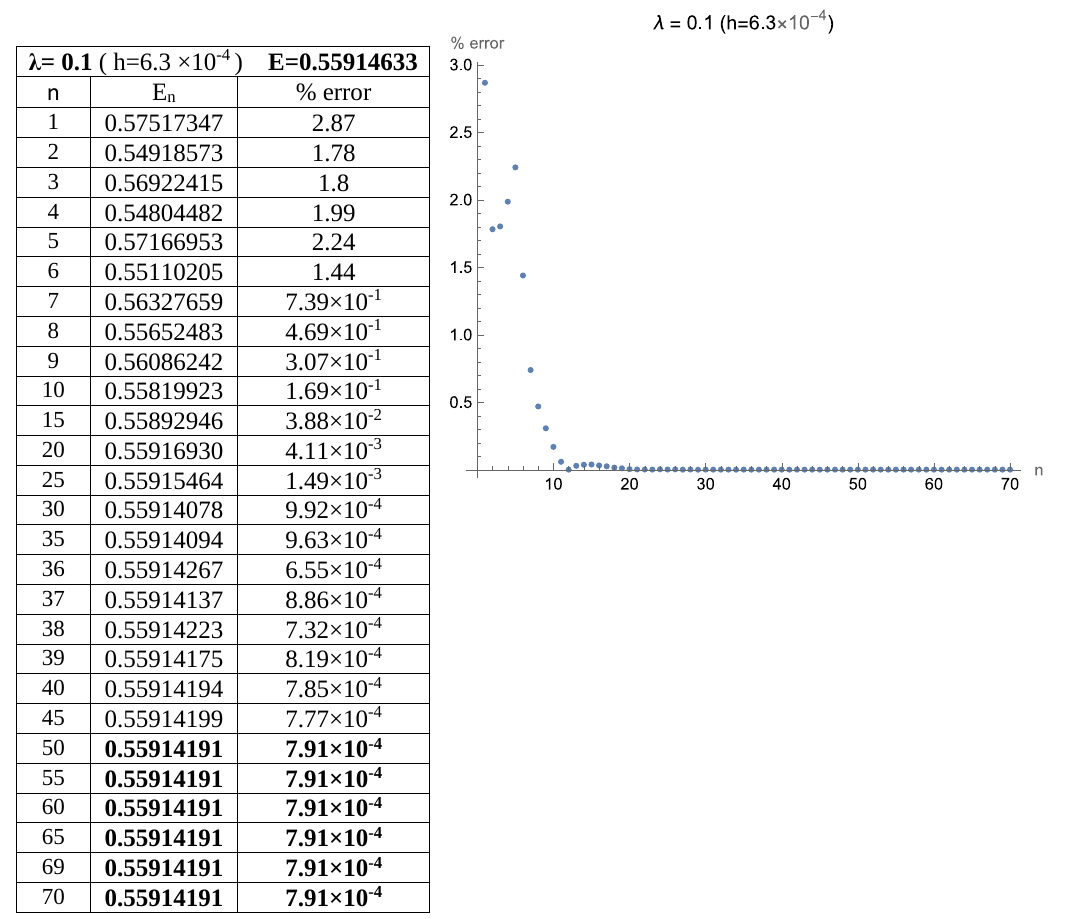}
		\caption{Results at intermediate coupling $\lambda=0.1$ with $h=6.3 \times 10^{-4}$. The parameter $h$ is smaller by a factor of $\approx 16$ compared to the previous case with $h=0.01$. The series now converges even closer to the exact value; the error is less than $8\times 10^{-4}\%$ which is extremely low and basically negligible.}     
		\label{Lambda001B}
\end{figure}

\begin{figure}[t]
	\centering
		\includegraphics[scale=0.8]{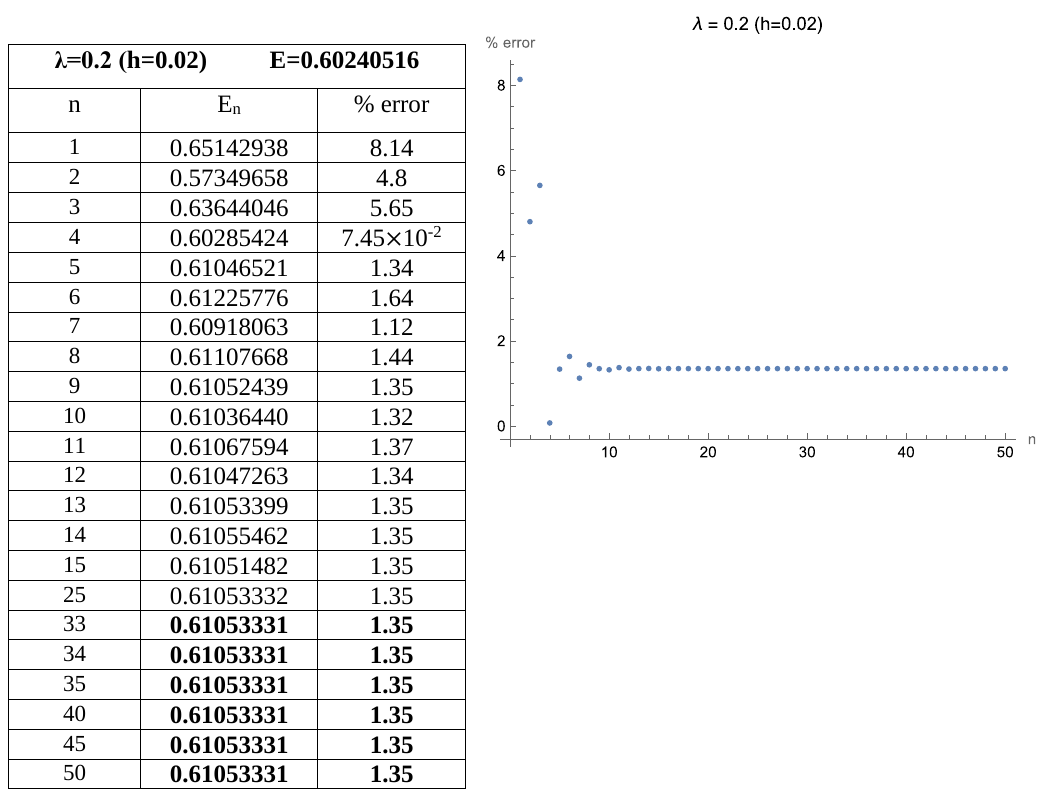}
		\caption{Results at strong coupling $\lambda=0.2$ for $h=0.02$. The series converges to the exact value to within $1.35 \%$ up to arbitrary large orders (shown here up to $n=50$). This is in stark contrast from the usual perturbative series which breaks down completely at this coupling; the series diverges right from the start (see Fig. \ref{Lambda022B}). We now have an absolutely convergent series that yields the correct answer to within a small error all the way to large orders.}     
		\label{Lambda22AA}
\end{figure}

\begin{figure}[t]
	\centering
		\includegraphics[scale=0.65]{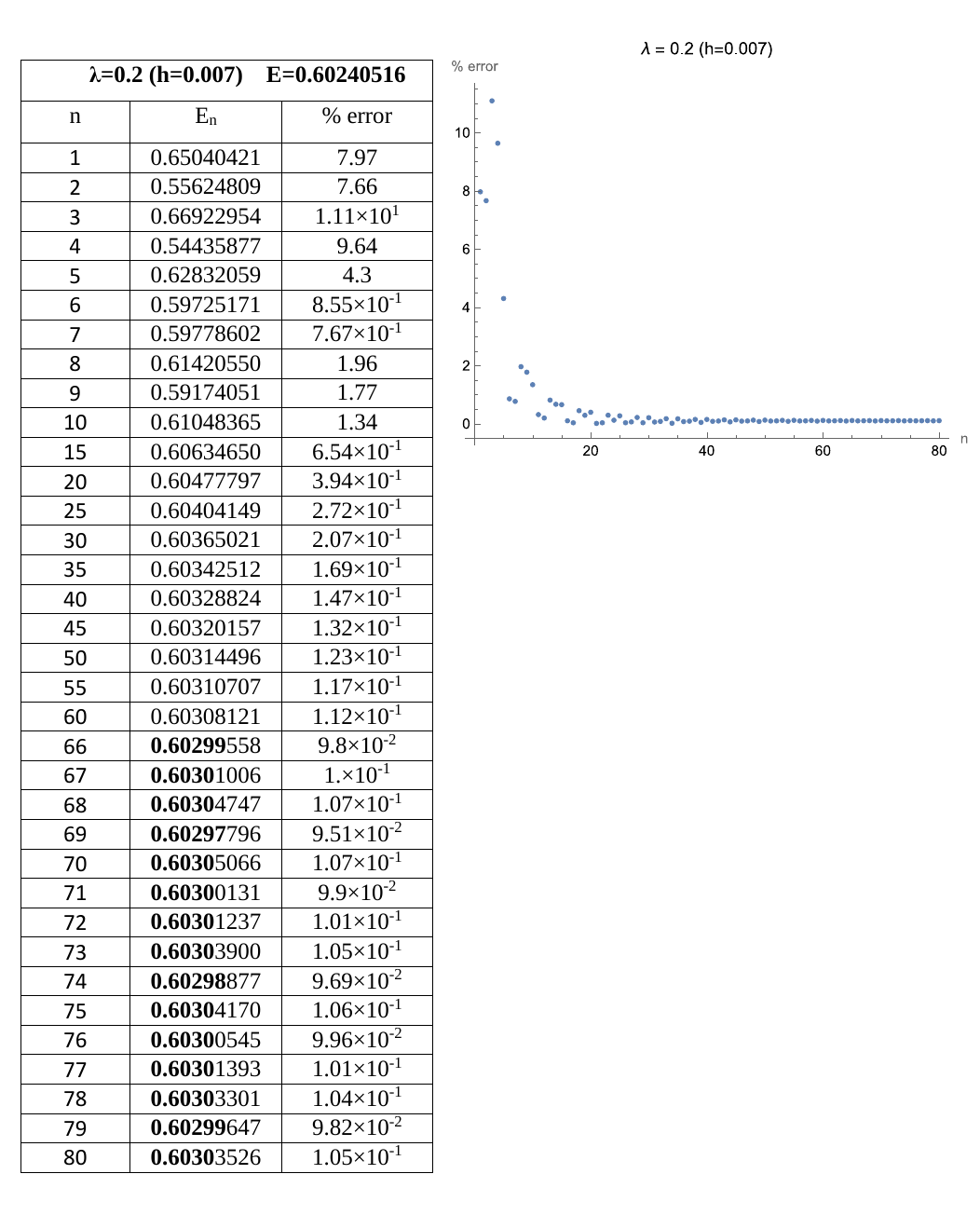}
		\caption{Results at strong coupling $\lambda=0.2$ for $h=0.007$. The parameter $h$ is smaller by a factor of around $5.8$ compared to the previous case of $h=0.02$. The plot shows that at this lower value of $h$ one needs to go to higher orders to reach convergence and that the value it converges to is closer to the exact value. The series now converges to within less than $0.11\%$ of the exact value (compared to $1.35\%$ at $h=0.02$). This is a remarkable result for a perturbative series at strong coupling. The series converges at four decimal places to $0.6030$ (shown in bold from $n=66$ to $n=80$) whereas the exact answer at four decimal places is $0.6024$ (a difference of $0.1\%$). Convergence of the series to more decimal places can be achieved by carrying out calculations at a higher level of numerical precision and going to higher orders}     
\label{Lambda22BB}
\end{figure}

Results are presented in figures \ref{Lambda002A} to \ref{Lambda22BB}. The figures contain a plot, a table of values and a description of the results for three different couplings: $\lambda=0.02$ (weak coupling), $\lambda=0.1$ (intermediate coupling), and $\lambda=0.2$ (strong coupling). For each coupling, we show results at two different values of the parameter $h$ where $h>0$ (the second value is the lowest and going lower did not change the results). In the table of values, the exact energy $E$ is placed at the top of the table and is quoted to eight decimal places. It is obtained by solving directly Schr\"odinger's equation numerically. The energy $E_n$ at order $n$ in the series is also quoted to eight decimal places and we calculate the $\%$ error between $E_n$ and the exact value $E$ at each order $n$. For each coupling at a given $h$, we plot the $\%$ error versus the order $n$. For a given coupling, the series converges closer to the exact energy for the smaller of the two values of $h$ as it should. The most important aspect of the results is that a perturbative series in powers of the coupling was used successfully to calculate the correct energy at strong coupling (to within a tiny error). This in turn is possible because the series expansion for the energy at a given parameter $h$ is an absolutely convergent series instead of an asymptotic one.  The results show that the series works extremely well at all three couplings: they converged to the exact energy to within a tiny $\%$ error.    




\subsection{Energy as a function of $h$ at fixed order $n_0$}
\begin{figure}
	\centering
	\begin{subfigure}[]{.48\textwidth}
		\includegraphics[scale=0.5]{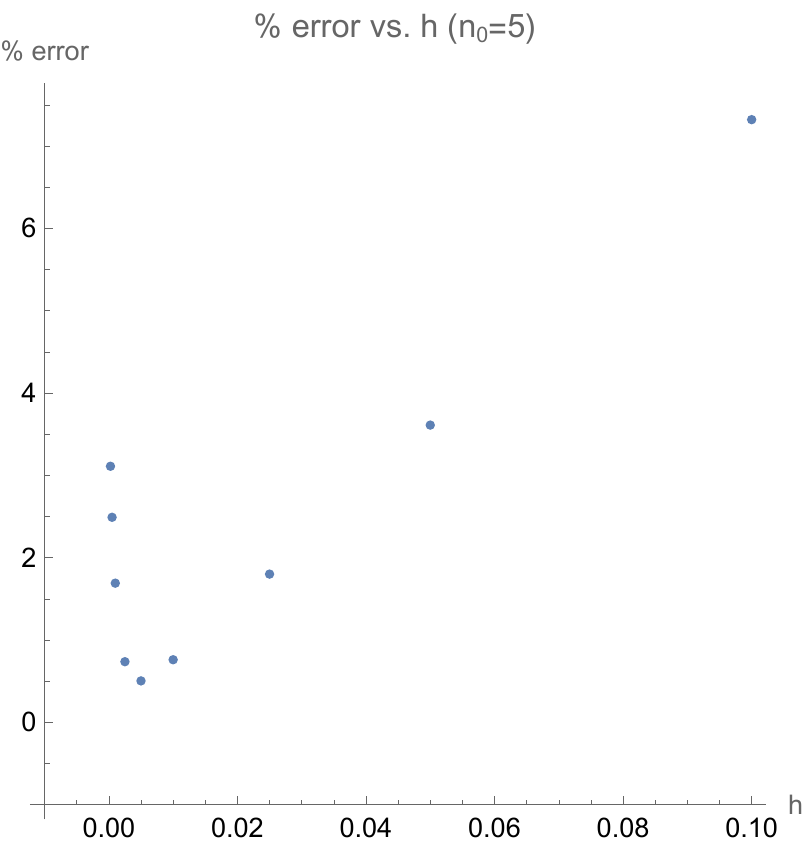}
		\caption{}     
		\label{Lambda0.1_n0=5}
\end{subfigure}
\hfill
\begin{subfigure}[]{.48\textwidth}
	\centering
		\includegraphics[scale=0.45]{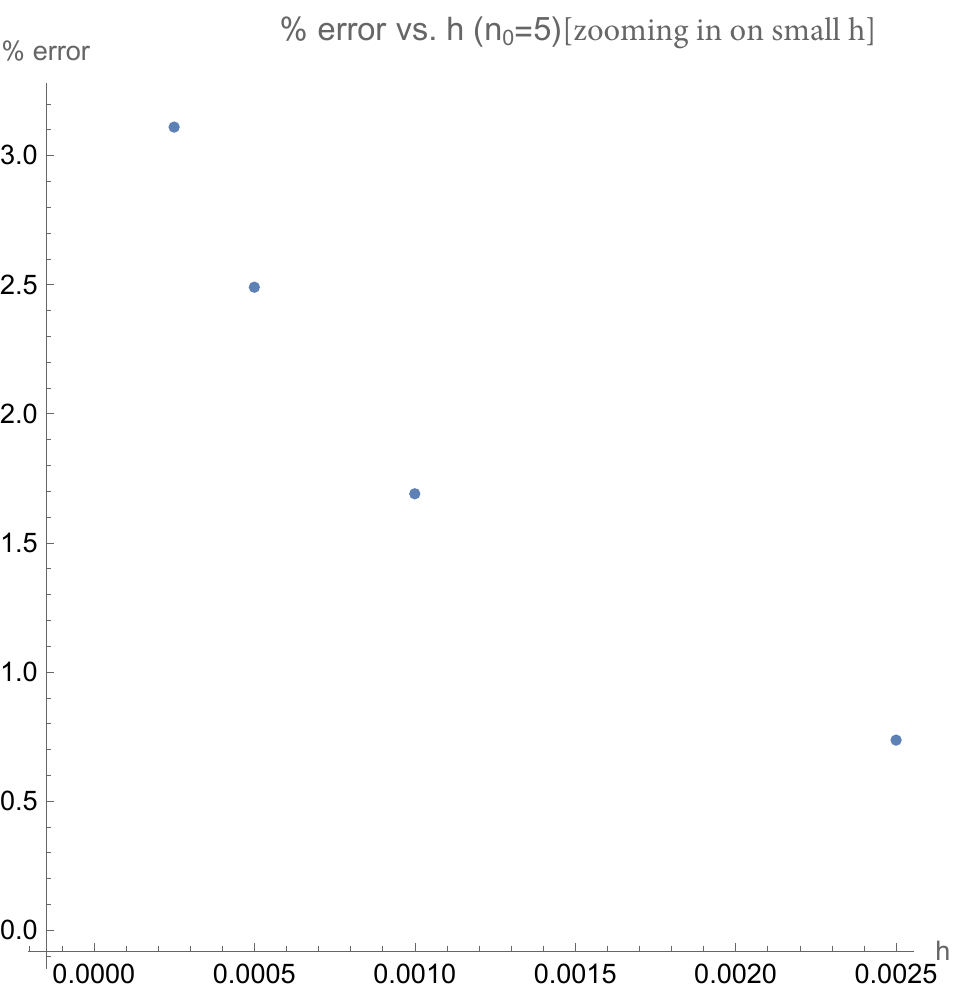}
		\caption{}      
		\label{Lambda0.1_n0=5_zoom}
\end{subfigure} 
\hfill
\begin{subfigure}[]{.88\textwidth}
	\centering
		\includegraphics[scale=0.70]{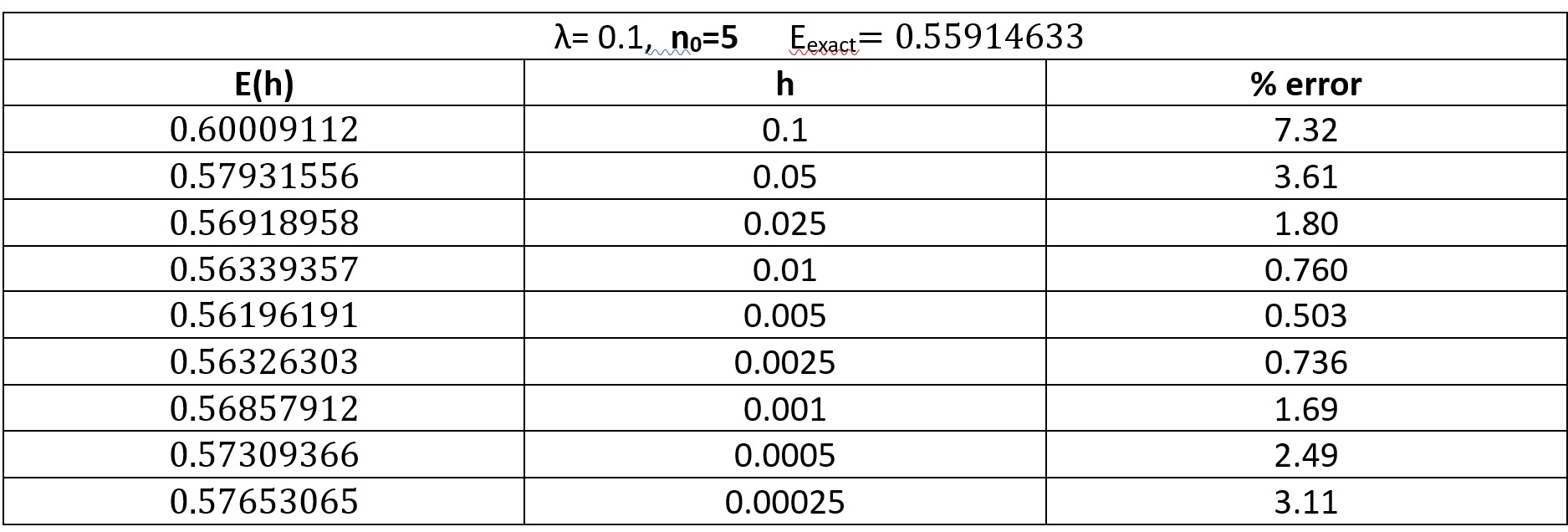}
		\caption{}      
		\label{Table_lambda0.1_n0=5}
\end{subfigure} 
\caption{$\%$ error versus $h$ for $n_0=5$ ($\lambda=0.1$). Note that the $\%$ error in Fig.(a) decreases as $h$ gets smaller except at the smallest values of $h$ where it is seen to increase (as seen clearly in Fig.(b) which zooms in on the smallest values of $h$). Table (c) contains the data plotted.}
\label{n05}
\end{figure}

\begin{figure}
	\centering
	\begin{subfigure}[]{.48\textwidth}
		\includegraphics[scale=0.45]{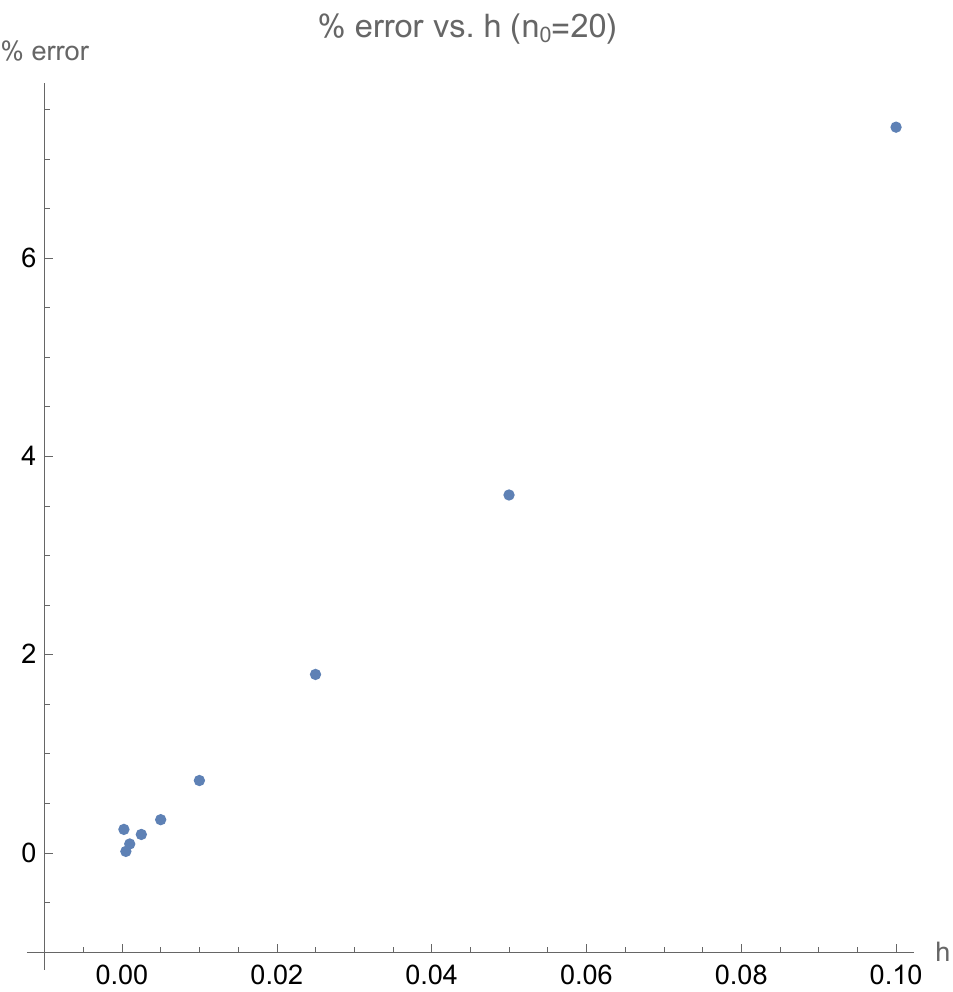}
		\caption{}     
		\label{Lambda0.1_n0=20}
\end{subfigure}
\hfill
\begin{subfigure}[]{.48\textwidth}
	\centering
		\includegraphics[scale=0.45]{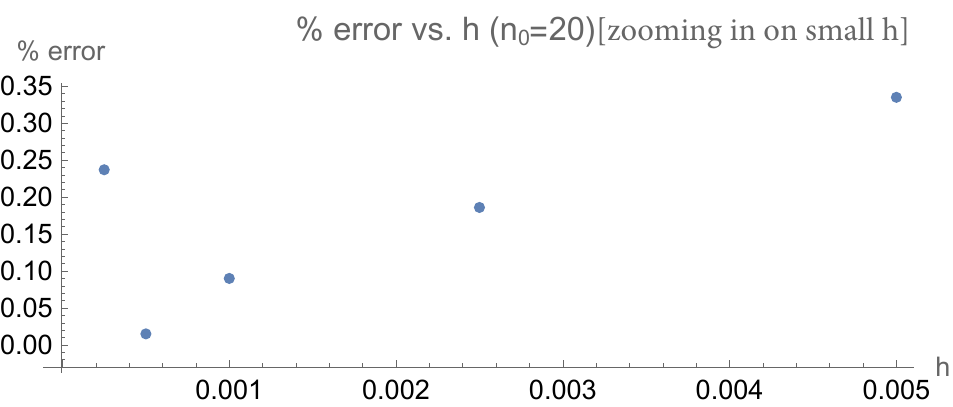}
		\caption{}      
		\label{Lambda0.1_n0=20_zoom}
\end{subfigure} 
\hfill
\begin{subfigure}[]{.88\textwidth}
	\centering
		\includegraphics[scale=0.70]{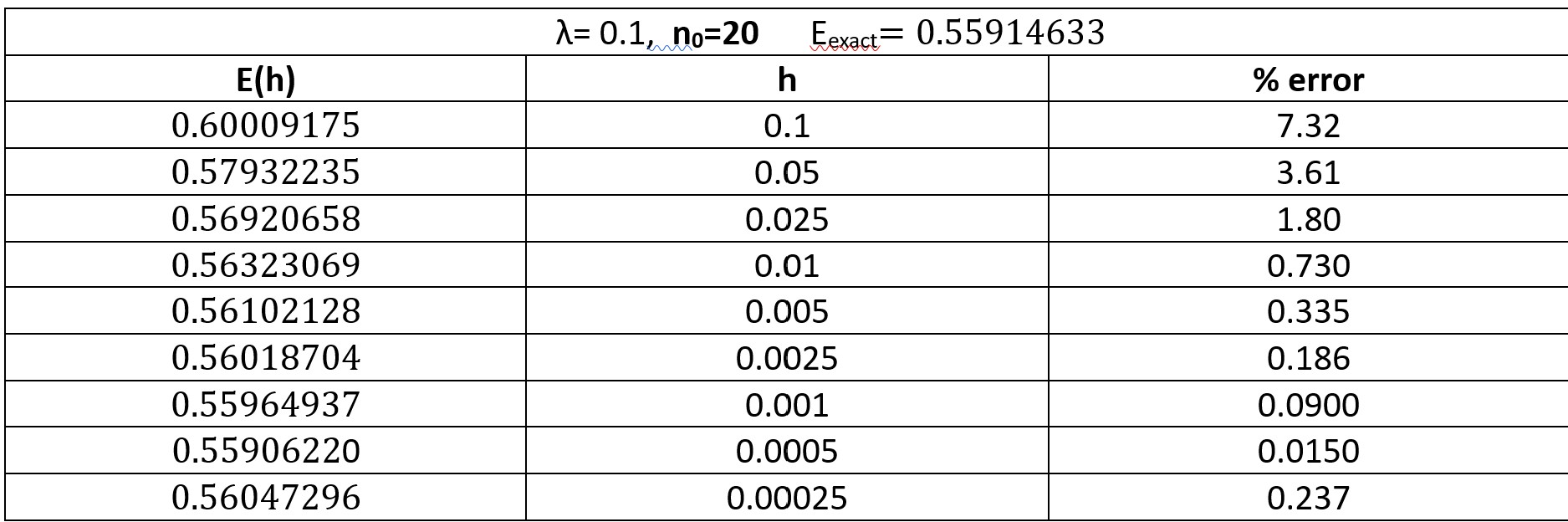}
		\caption{}      
		\label{Table_lambda0.1_n0=20}
\end{subfigure} 
\caption{$\%$ error versus $h$ for $n_0=20$ ($\lambda=0.1$). Note that the $\%$ error in Fig. (a) decreases as $h$ gets smaller except at the smallest value of $h$ where it is seen to increase (as seen clearly in Fig.(b) which zooms in on the smallest values of $h$). Table (c) contains the data plotted.}
\label{n020}
\end{figure}

\begin{figure}
	\centering
	\begin{subfigure}[]{.48\textwidth}
		\includegraphics[scale=0.45]{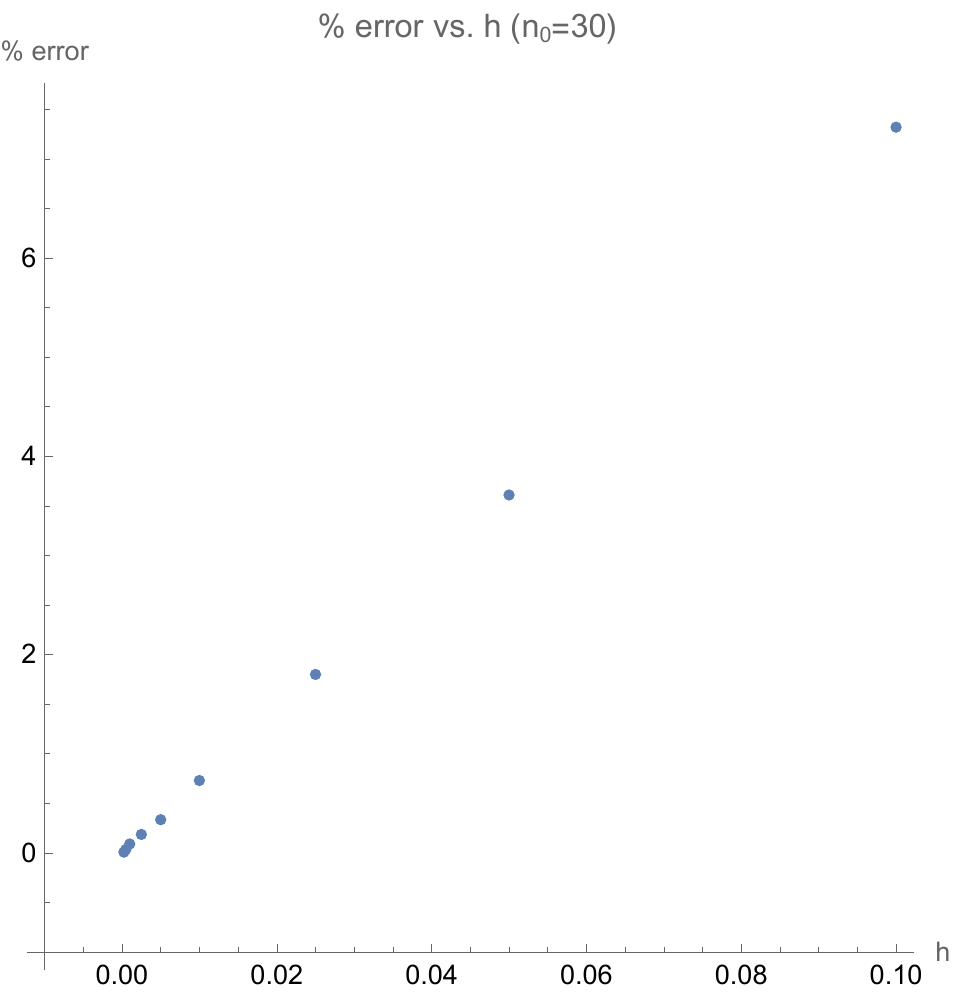}
		\caption{}     
		\label{Lambda0.1_n0=30}
\end{subfigure}
\hfill
\begin{subfigure}[]{.48\textwidth}
	\centering
		\includegraphics[scale=0.45]{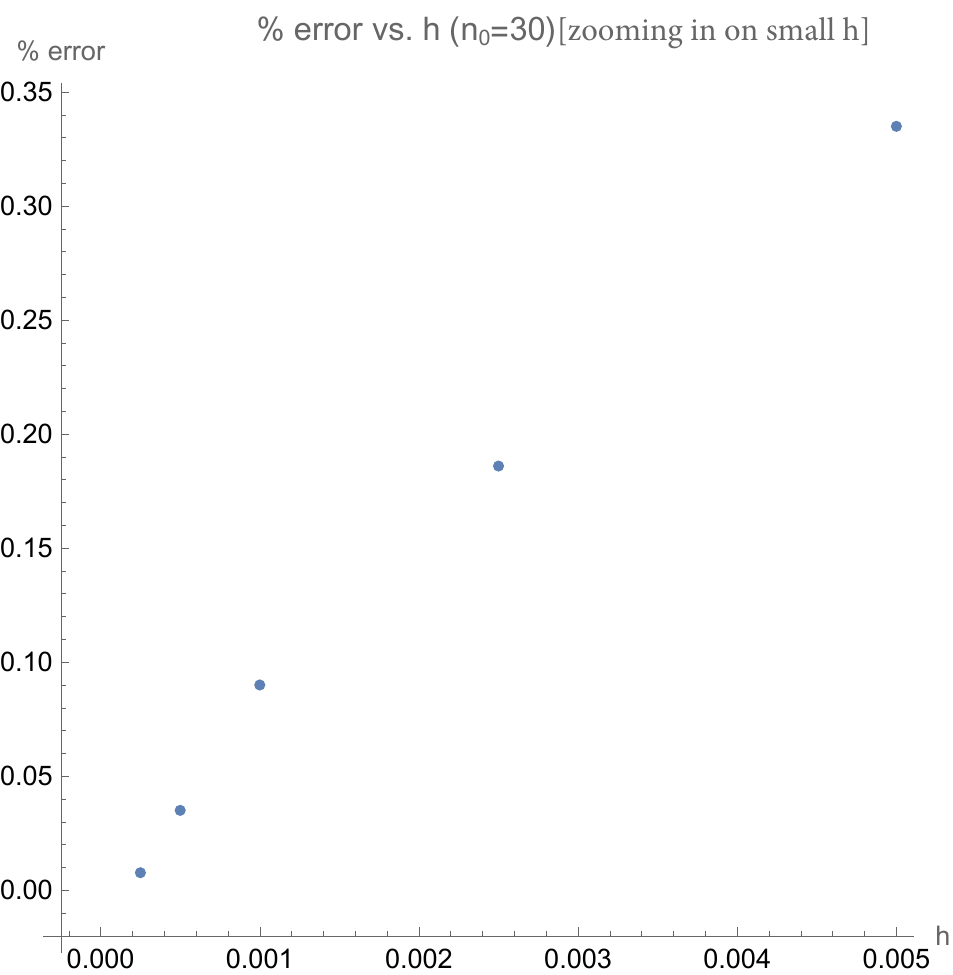}
		\caption{}      
		\label{Lambda0.1_n0=30_zoom}
\end{subfigure} 
\hfill
\begin{subfigure}[]{.88\textwidth}
	\centering
		\includegraphics[scale=0.60]{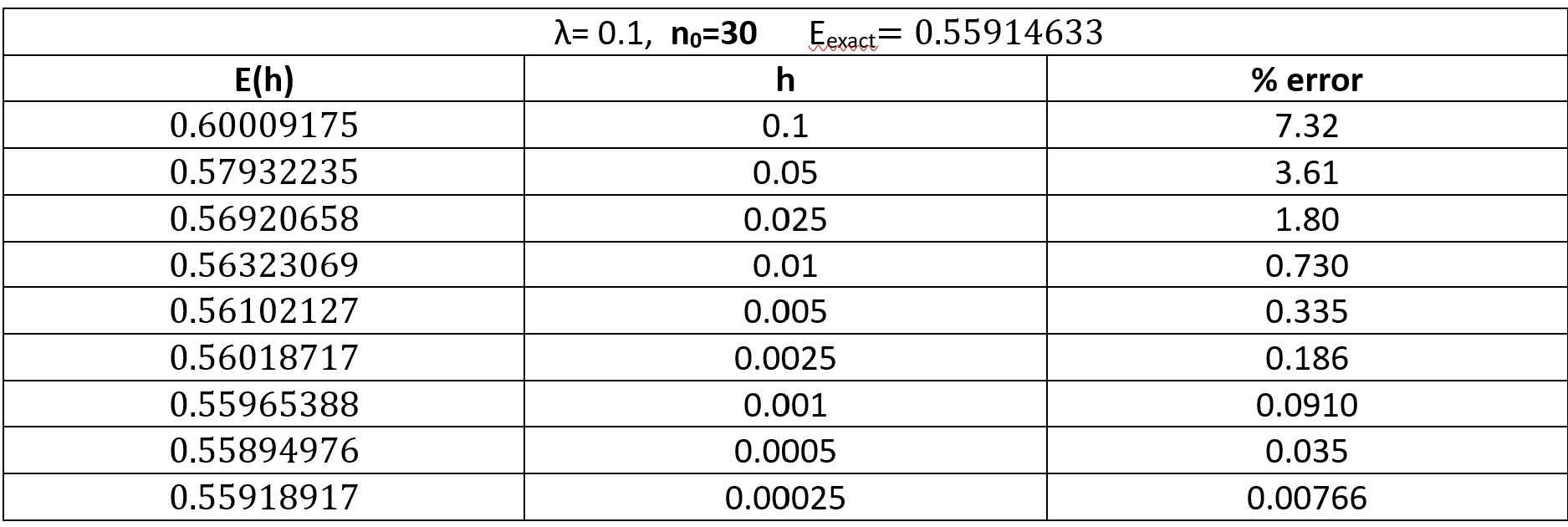}
		\caption{}      
		\label{Table_lambda0.1_n0=30}
\end{subfigure} 
\caption{$\%$ error versus $h$ for $n_0=30$ ($\lambda=0.1$). In contrast to $n_0=5$ and $n_0=20$, the $\%$ error in Fig.(a) decreases monotonically all the way to the smallest values of $h$. This is seen even more clearly in Fig.(b). Table (c) contains the data plotted.}
\label{n030}
\end{figure}

In figures \ref{Lambda002A} to \ref{Lambda22BB} we saw how the energy varies with order $n$ at a fixed value of $h$ and $\lambda$. We did this for two different values of $h$ and three different values of $\lambda$. However, it is also of significant interest to see how the energy varies with $h$ at a fixed order $n$ which we label $n_0$ \footnote{The author thanks the referee for suggesting this study.}. We considered three different values of $n_0$ with $\lambda$ fixed at its intermediate value of $\lambda=0.1$ (this is a good value to work with since regular perturbation theory diverges early on at this coupling). For each $n_0$, we plot the energy as a function of $h$. The plot shows that the error in the energy decreases monotonically as $h$ decreases but that this trend does not continue all the way to the smallest values of $h$ in our sample. To see this clearly, we made two plots for each $n_0$: one that covers the entire range of $h$ in our sample and one that zooms in at small $h$. We also provide a table of values for each $n_0$. So each figure contains two plots and one table for each $n_0$. These results are presented in figures \ref{n05}, \ref{n020} and \ref{n030} corresponding to $n_0=5$, $n_0=20$ and $n_0=30$ respectively. The most important observation is that $n_0$ has to be made sufficiently large (a lower bound) in order for the error in the energy to decrease continuously as $h$ approaches its smallest values. For our sample, this occurred roughly at $n_0=30$ (see fig. \ref{n030}). For the cases where $n_0$ was smaller than this, the error in the energy decreases as $h$ decreases but starts to increase at the lowest values of $h$ (see figures \ref{n05} and \ref{n020}).   

\section{The sextic anharmonic oscillator} 

As a further application of our method based on finite path integral limits, we now consider the sextic anharmonic potential. This potential finds many applications in various fields in physics including the study of molecular vibrations as well as in condensed matter physics. The goal is to obtain the ground state energy of the potential $V(x)= x^2/2 + \lambda\,x^6$ via a convergent perturbative series that is valid both at weak and strong coupling $\lambda$. We do not include in the potential a quartic part in order to make a clear comparison between the quartic and sextic case.   We will first consider the usual (asymptotic) perturbative series. We will see again that there is a weak coupling regime where the series has a plateau region at a value close to the correct energy before it ultimately diverges. However, at higher couplings, which we label intermediate and strong, the perturbative series is unreliable as it never settles (plateaus) to any value and diverges quickly. Strong coupling now occurs at a much lower value of the coupling $\lambda$ compared to the quartic case, so that the usual perturbative series breaks down earlier. We then obtain a convergent perturbative series based on finite path integral limits; this becomes even more important now as it is needed at lower values of the coupling $\lambda$ than in the quartic case.

\subsection{The usual (asymptotic) perturbative series} 
We will obtain a recursion relation for the sextic anharmonic potential that is the analog of the recursion relation \reff{Recursion} of the quartic case. This yields the coefficients that enter the usual perturbative series. We will naturally be more brief here as the procedure is similar to that presented at the end of section 3 for the quartic case. We begin with Schr\"odinger's equation for the sextic anharmonic oscillator which is (setting $\hbar=m=\omega=1$):
\beq
\Big(-\dfrac{1}{2} \dfrac{d^2}{dx^2} +\dfrac{x^2}{2} + \lambda\, x^6\Big) \psi(x)= E_0\,\psi(x)\,.
\eeq{Schro3}
We express again the ground state energy $E_0$ and  wavefunction $\psi(x)$ as a power series in $\lambda$:
\begin{align}
E_0=\frac{1}{2} +\sum_{j=1}^{\infty} a_j \, \lambda^j
\label{E03}
\end{align}
and   
\beq
\psi(x)= e^{-x^2/2}\sum_{m=0}^{\infty} \lambda^m\,B_m(x)
\eeq{Ansatz3}
where $e^{-x^2/2}$ is the harmonic oscillator wavefunction (up to a normalizing factor) and $B_m(x)$ are polynomial functions with $B_0(x)=1$. The coefficients $a_j$ will be obtained via a recursion relation. Substituting \reff{Ansatz3} and \reff{E03} into \reff{Schro3} we obtain 
\beq
x\,B_n^{'}(x) -\dfrac{1}{2}\, B_n^{''}(x) + x^6 \,B_{n-1}(x) =\sum_{p=0}^{n-1} a_{n-p} B_p(x)\,.
\eeq{RecurF3}
The functions $B_n(x)$ are now polynomials of degree $6n$ (compared to $4n$ for the quartic case) containing only even powers of $x$ :
\beq
B_m(x)=\sum_{j=0}^{3 m} (-1)^m x^{2\,j}\, B_{m,j} 
\eeq{RecurB3}     
where $B_{0,0}=1$ and $B_{k,0}=0$ for $k\ne 0$ so that $B_0(x)=1$ and 
$B_m(x)$ contains no constants for $m\ne 0$. 

By matching a constant on both sides of \reff{RecurF3} we obtain the same relation as \reff{Bn1}:  
\beq
a_n=(-1)^{n+1}\,B_{n,1}\,.
\eeq{Bn12}
Substituting \reff{RecurB3} and \reff{Bn12} into the equation \reff{RecurF3} we obtain the recursion relation for the coefficients $B_{i,j}$:
\beq
2\,j B_{i,j}- (j+1) (2j+1) B_{i, j+1} - B_{i-1,j-3}= - \sum_{p=1}^{i-1} B_{i-p,1}\,B_{p,j}\,.
\eeq{Recursion3}
Note that we have $B_{i-1,j-3}$ above in contrast to $B_{i-1,j-2}$ for the quartic case. In the above $i \ge 1$ and $j \ge 1$. We have that $B_{0,0}=1$, $B_{k,0}=0$ for $k \ne 0$ and $B_{m,k}=0$ if $k>3\,m$ or $k<0$. We generated the coefficients $B_{n,1}$ from $n=1$ to $n=50$ i.e. the first $50$ coefficients $a_n$ that enter the energy expansion \reff{E03}.   

We obtained results at three separate couplings: weak ($\lambda=0.0005$), intermediate ($\lambda=0.005$) and strong ($\lambda=0.02$). Intermediate coupling, by definition, is where the perturbative series may have one or two good points at small order $n$ but is ultimately unreliable as it does not plateau (settle) to the correct value before diverging in contrast to weak coupling. Note that this now occurs for the sextic case at $\lambda=0.005$ which is lower by a factor of $20$ compared to the value of 
$\lambda=0.1$ for the quartic case. Strong coupling is where perturbation theory breaks down completely and diverges quickly right from the start. This now occurs for the sextic case at $\lambda=0.02$ which is lower by a factor of $10$ compared to the value of $\lambda=0.2$ for the quartic case. These are rough estimates but show clearly that \textit{perturbation theory becomes unreliable at significantly lower couplings $\lambda$ in the sextic case compared to the quartic case}. Therefore, the method of finite path integral limits that was used to obtain a reliable convergent perturbative series in the quartic case becomes even more important in the sextic case. 

\begin{figure}[t]
	\centering
		\includegraphics[scale=0.8]{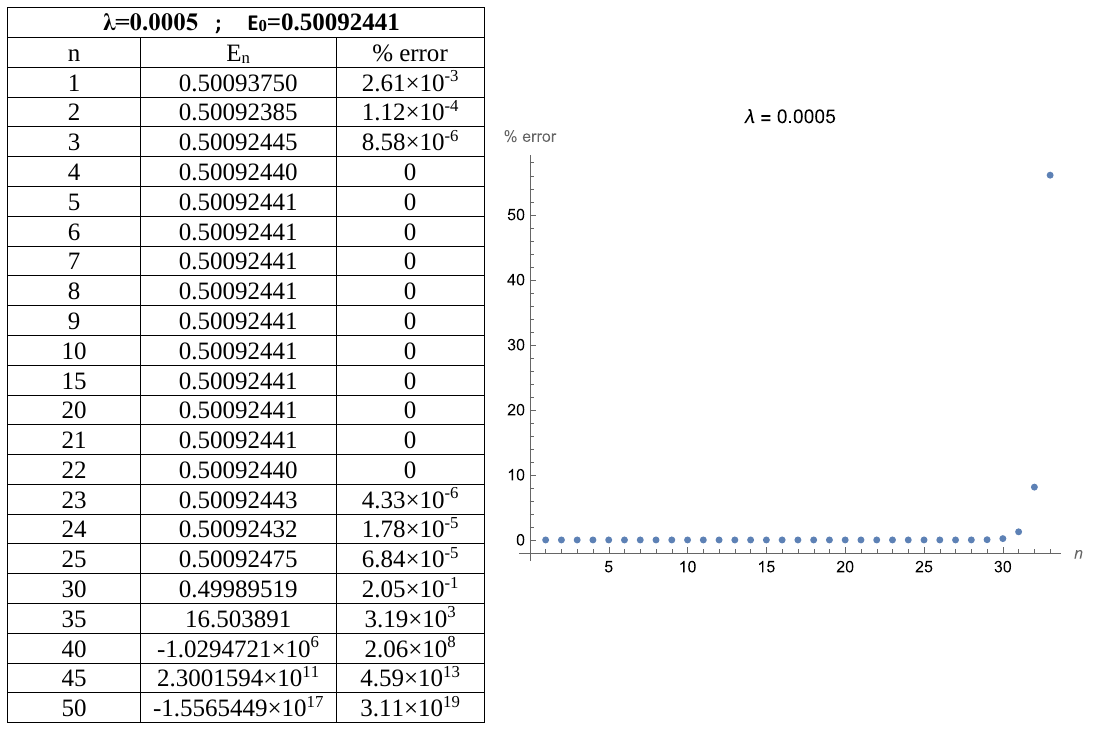}
		\caption{Results at weak coupling $\lambda=0.0005$. The table contains the $\%$ error between the energy $E_n$ and the exact ground state energy $E_0=0.50092441$ obtained numerically. Both $E_n$ and $E_0$ are quoted to eight decimal places. The $\%$ error is completely negligible over a long range of orders (from roughly $n=1$ to $n=30$)and is in fact zero from $n=4$ to $n=22$ (to within eight decimal accuracy). The long plateau region on the plot implies that the series can be used to make reliable predictions. After the plateau region, the series begins to diverge and reaches a very large $\%$ error of order $10^19$ at $n=50$. It is an asymptotic series.}     
		\label{WeakSext}
\end{figure}

\begin{figure}[t]
	\centering
		\includegraphics[scale=0.8]{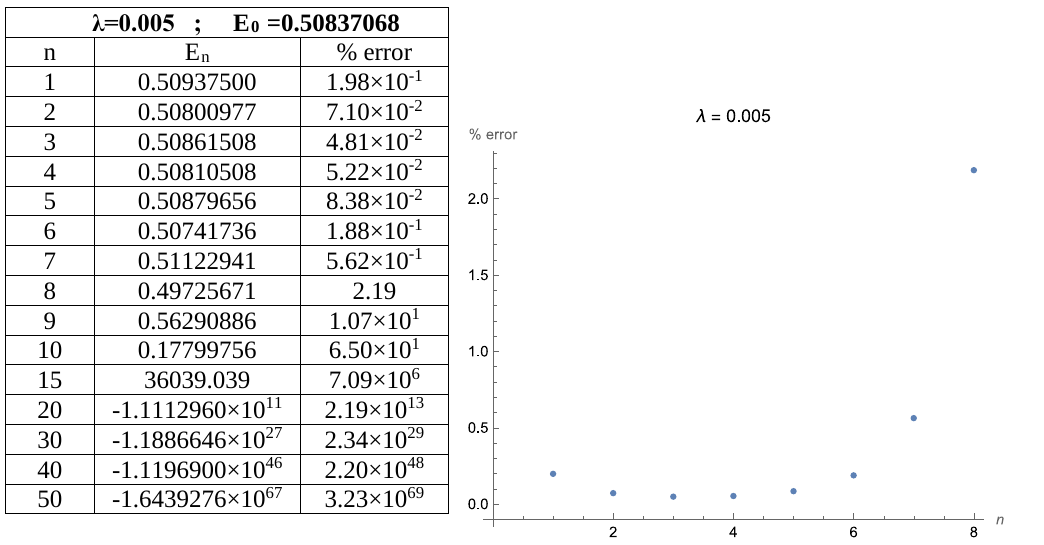}
		\caption{Results at intermediate coupling $\lambda=0.005$. The $\%$ error between the series $E_n$ and the exact value $E_0=0.50837068$ starts at a low value of $0.2\%$, and dips to a minimum of $0.05\%$ at $n=3$ and then quickly diverges after $n=8$(asymptotic series). So the series is close to the exact value early on (at low orders) but in contrast to the weak coupling case, there is no plateau region where it settles close to the exact value over a long range of orders.}     
		\label{IntSext}
\end{figure}

\begin{figure}[t]
	\centering
		\includegraphics[scale=0.8]{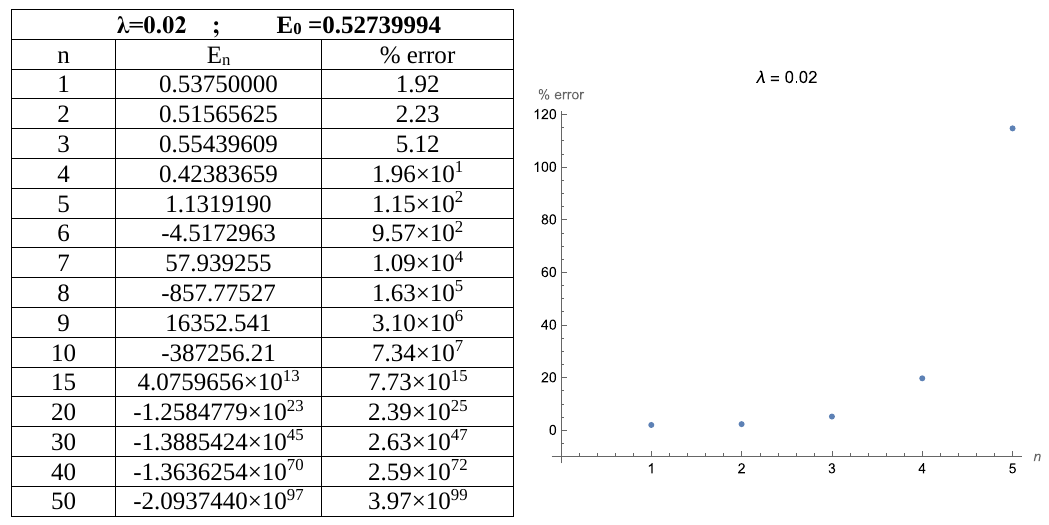}
		\caption{Results at strong coupling $\lambda=0.02$. The $\%$ error starts at $1.98\%$and increases monotonically afterwards. At $n=5$ the error reaches already $115\%$ and at $n=50$ the error is an astronomical $10^{99}\%$. At strong coupling, the perturbative series breaks down completely. This is in strark contrast to the absolutely convergent perturbative series of the next section which yields excellent results at strong coupling (less than $1\%$ error)}     
		\label{StrongSext}
\end{figure}

For each coupling, we calculated the energy $E_n$ at each order $n$ using \reff{E03} from $n=1$ to $n=50$.  We also calculate the $\%$ error between $E_n$ at a given order and the exact value $E_0$ for the ground state energy obtained by solving numerically Schr\"odinger's equation for the sextic anharmonic potential. The results are presented in table form and in a plot for each coupling in figures \ref{WeakSext}, \ref{IntSext} and \ref{StrongSext}. The plot in fig. \ref{WeakSext} for weak coupling ($\lambda=0.0005$) shows a long plateau region where the series settles to the correct value (to within $0.3 \%$) before it begins to diverge at larger orders (starting roughly at $n>30$). The series is an asymptotic series but is still useful since one can extract from it correct results over a long range of orders. The plot in fig. \ref{IntSext} for intermediate coupling ($\lambda=0.005$) dips close to the exact value after a few orders ($\%$ error reaches a minimum of $0.05 \%$ at $n=3$) but then diverges shortly afterwards. In contrast to weak coupling, there is no plateau region where it settles close to the exact value over a long range of orders. The plot in fig. \ref{StrongSext} for strong coupling ($\lambda=0.02$) shows the $\%$ error increasing monotonically with order from the start. The series diverges very rapidly (reaches huge errors at small order) so that the perturbative series breaks down completely. The divergence at strong coupling is significantly worse in the sextic than in the quartic case. This renders the absolutely convergent perturbative series developed for the sextic case all the more important.  


\subsection{Convergent perturbative series for the energy of the sextic anharmonic oscillator at strong coupling}

As in the quartic case we now place infinite walls at $x=\pm L$ in the sextic anharmonic potential where $L$ is an arbitrary finite length (we \textit{formally} recover the original potential in the limit $L\to \infty$). Without walls the wavefunction for the harmonic potential is given by $e^{-x^2/2}$ and with infinite walls it is given by \reff{WallsHO} (see section 4 for derivation):  
\beq
\psi_0(x) =e^{-x^2/2}\,{}_1F_1\big(-\frac{h}{4}\,;\,\frac{1}{2}\,; \,x^2\,\big)\,.
\eeq{WallsHO2}
where ${}_1F_1\big(-\frac{h}{4}\,;\,\frac{1}{2}\,; \,x^2\,\big)$ is the Kummer confluent hypergeometric function. It is an even function of $x$ and the parameter $h$ is positive and its value determines $L$. At $h=0$, the function is unity and this is equivalent to having no walls (in effect $L\to \infty$). As $h$ increases, $L$ decreases. Figure 4 shows a plot of the function for $h=0.001$.

The ground state energy of the harmonic oscillator as a function of $h$ was already obtained in section 4 and is given by \reff{Eh}:
\beq
E_{\lambda=0}= \dfrac{1+h}{2}\,.
\eeq{Eh2}
. We can now express the ground state wavefunction $\psi(x)$ of the sextic anharmonic oscillator with walls as a series expansion in powers of $\lambda$ with the zeroth order being the harmonic solution \reff{WallsHO2}:
\beq
\psi(x)=\psi_0(x)\,\sum_{n=0}^{\infty} \lambda^n\,B_n(x)= e^{-x^2/2}\,{}_1F_1\big(-\frac{h}{4}\,;\,\frac{1}{2}\,;\, x^2\,\big)\,\sum_{n=0}^{\infty} \lambda^n\,B_n(x)
\eeq{Ansatz4} 
where $B_n(x)$ are again polynomials in even powers of $x$ with $B_0(x)=1$. Since $\psi_0(x)$ is zero at the location of the walls ($x=\pm L$), $\psi(x)$ is guaranteed to be zero at the walls also (where $L$ is determined by $h$). The ground state energy of the sextic anharmonic oscillator with walls is now a function of $h$ and can be expressed as a series expansion in powers of $\lambda$:
\beq
E_0= \dfrac{1+h}{2} +\sum_{k=1}^{\infty} a_k \, g^k 
\eeq{E0h2}
where we used the result \reff{Eh2}. The coefficients $a_k$ depend on $h$. We recover the coefficients and energy of section 5.1 (no walls) in the limit $h\to 0$. We now require a new recursion relation to calculate the coefficients $a_k$ for a given $h$. This will be a modified version of our previous recursion relation \reff{Recursion3}. Schr\"odinger's equation for the sextic anharmonic oscillator is given by \reff{Schro3}: 
\beq
\Big(-\dfrac{1}{2} \dfrac{d^2}{dx^2} +\dfrac{x^2}{2} + \lambda\, x^6\Big) \psi(x)= E_0\,\psi(x)\,.
\eeq{SchroA3}
Substituting \reff{Ansatz4} and \reff{E0h2} into \reff{SchroA3} and matching powers of 
$\lambda$ on both sides we obtain the following differential equation:
\begin{align}
(2+h) \,x\,B_n'(x)\,F_2 + F_1\,\big(- x \,B_n'(x)-\frac{B_n''(x)}{2}+\,x^6 \,B_{n-1}(x)\,\big)= F_1\,\sum_{p=0}^{n-1} a_{n-p}\,B_p(x) \,.
\label{Dif3}  
\end{align} 
where
\begin{align}
F_1&= \,_1F_1\big(-\frac{h}{4}\,;\,\frac{1}{2}\,;\,x^2\,\big)\nonumber \\
\text{and}\nonumber\\
F_2&=\, _1F_1\big(-\frac{h}{4}\,;\,\frac{3}{2}\,;\,x^2\,\big)\,.
\label{FF12}
\end{align}
As in section 5.1, we write $B_n(x)$ as polynomials of degree $6n$ containing only even powers of $x$ :
\beq
B_n(x)=\sum_{j=0}^{3 n} (-1)^n x^{2\,j}\, B_{n,j}\,.
\eeq{BPoly2}     
Again, we define $B_{0,0}=1$ and $B_{k,0}=0$ for $k\ne 0$ so that $B_0(x)=1$ and $B_m(x)$contains no constants for $m\ne 0$. The hypergeometric functions $F_1$ and $F_2$ defined by \reff{FF12} have the following series expansions:
\begin{align}
F_1&=\sum_{m=0}^{\infty} c_m \dfrac{(x^2)^m}{m!}=1+ c_1\, x^2 +\frac{1}{2!} \,c_2\,x^4 +...
\nonumber\\
F_2&=\sum_{m=0}^{\infty} b_m \dfrac{(x^2)^m}{m!}==1+ b_1\, x^2 +\frac{1}{2!}\, b_2\,x^4 +...
\label{Fifa2}
\end{align}
where $c_m$ and $b_m$ are coefficients given by \reff{cb}:
\begin{align}
c_m&=\dfrac{\Gamma(m-h/4)\, \Gamma(1/2)}{\Gamma(-h/4) \,\Gamma(m+1/2)}\nonumber\\
b_m&=\dfrac{\Gamma(m-h/4)\, \Gamma(3/2)}{\Gamma(-h/4) \,\Gamma(m+3/2)}\,.
\label{cb2}
\end{align}
We have $c_0=b_0=1$ regardless of the value of $h$. Also, $\lim_{h\to 0} b_m= \lim_{h\to 0} c_m =0$ for $m \ge 1$. The equation \reff{Dif3} yields again 
\beq
a_n= (-1)^{n+1}\, B_{n,1}\,.  
\eeq{an4}
Substituting \reff{an4}, \reff{BPoly2} and \reff{Fifa2} into \reff{Dif3} we obtain our new recursion relation:
\pagebreak
\begin{align}
\sum_{\substack{j=1\\m=k-j\\ 1\le k \le 3n}}^{k}& \Big[2\,j B_{n,j}\Big((2+h)\,\frac{b_m}{m!}-\frac{c_m}{m!}\Big) -(j+1)\,(2j+1) B_{n,j+1} \frac{c_m}{m!}\nonumber\\
&\qquad- B_{n-1,j-3}\,\frac{c_m}{m!}+\sum_{p=1}^{n-1} B_{n-p,1}\,B_{p,j}\,\frac{c_m}{m!}\Big]=0\,.
\label{Recursionh2}
\end{align}
The parameter $h$ appears in \reff{Recursionh2} both explicitly and via the coefficients $b_m$ and $c_m$ given by \reff{cb2}. The above recursion relation yields one equation for each $k$ where $k$ runs from $1$ to $3 n$. At each $k$, the equation is composed of a sum of terms from $j=1$ to $k$ with $m=k-j$. At each order $n$, one can then solve for the coefficient $B_{n,1}$ (and hence $a_n$) for a given $h$ by solving sequentially the set of $3n$ equations.  

The series for the energy expansion \reff{E0h2} is now an \textit{absolutely convergent} series for any finite positive value of $h$ ($h\ne 0$). When the energy to a desired level of accuracy no longer changes as $h$ decreases one is then close to the exact energy (i.e. close to within that accuracy). 

\begin{figure}[t]
	\centering
		\includegraphics[scale=0.8]{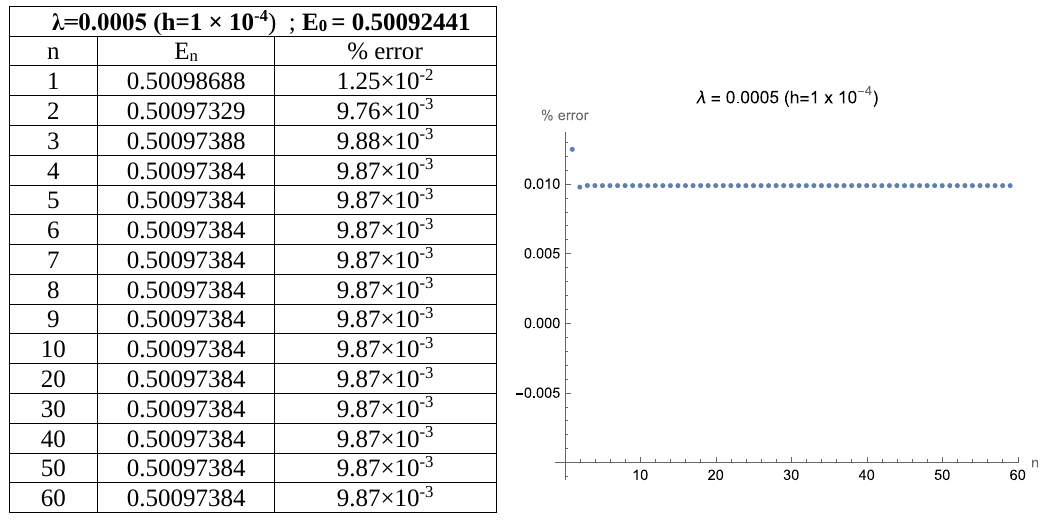}
		\caption{Results at weak coupling $\lambda=0.0005$ with $h=1 \times 10^{-4}$. The series converges to the exact value of the energy to within less than $0.01\%$ up to arbitrary large orders (we show it here up to $n=60$). It is an absolutely convergent series. This is in contrast to the usual perturbative series at weak coupling which diverges at large order as in the plot of Fig. \ref{WeakSext}.}     
		\label{Weakh1Sextic}
\end{figure}

\begin{figure}[t]
	\centering
		\includegraphics[scale=0.8]{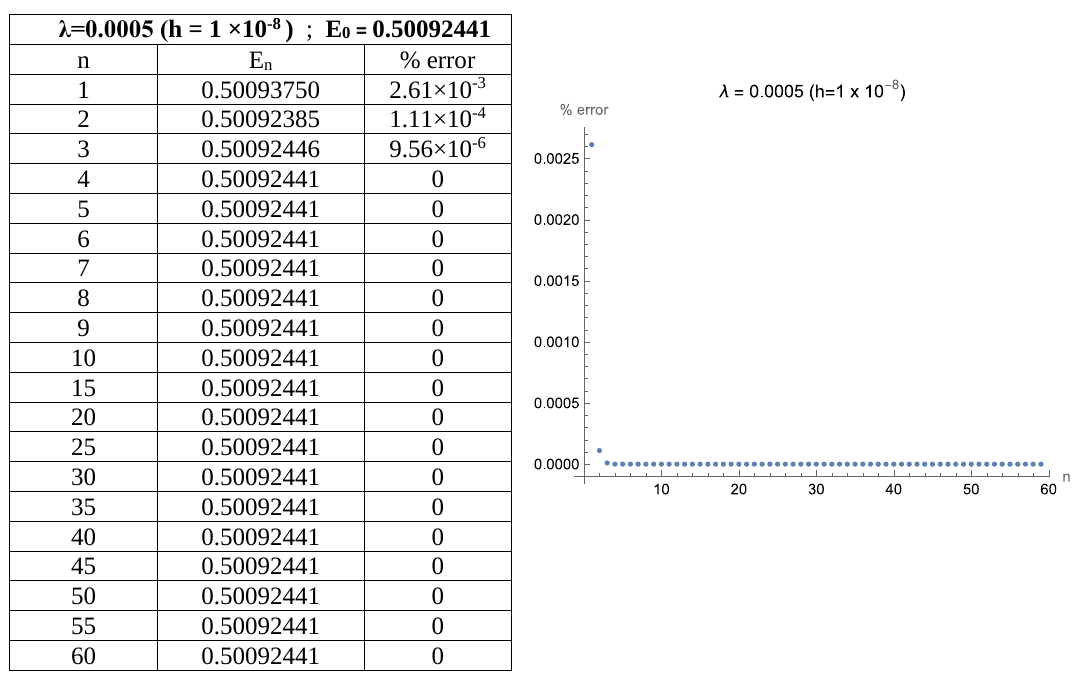}
		\caption{Results at weak coupling $\lambda=0.0005$ with $h=1\times 10^{-8}$. The parameter $h$ is smaller by a factor of $10^4$ compared to the previous case with $h=1\times 10^{-4}$. The series now matches the correct value to eight decimal accuracy (hence zero error to within eight decimal places). It converges up to arbitrary large orders (shown here up to $n=60$). Again this is an absolutely convergent series instead of an asymptotic one.}      
		\label{Weakh2Sextic}
\end{figure}

\begin{figure}[t]
	\centering
		\includegraphics[scale=0.8]{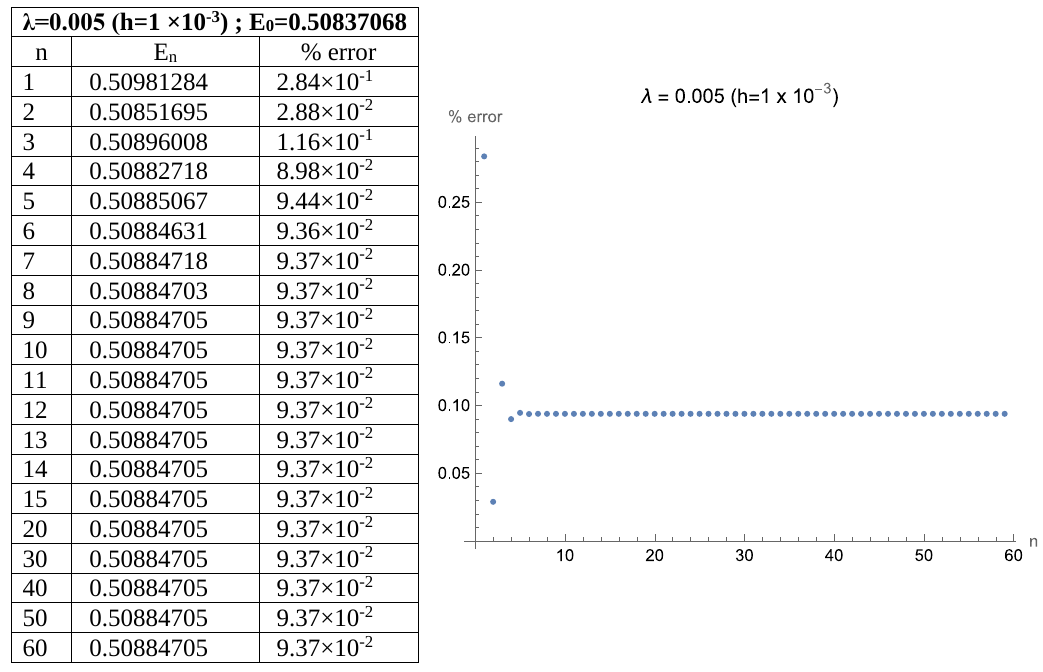}
		\caption{Results at intermediate coupling $\lambda=0.005$ with $h=0.001$. The series converges to the exact value of the energy to within less than $0.1\%$ up to arbitrary large orders (we show it here up to $n=60$). It is an absolutely convergent series. In contrast, the usual perturbative series at this coupling began to diverge at low orders after an initial dip (see Fig.\ref{IntSext}). }     
		\label{Inth1Sextic}
\end{figure}

\begin{figure}[t]
	\centering
		\includegraphics[scale=0.8]{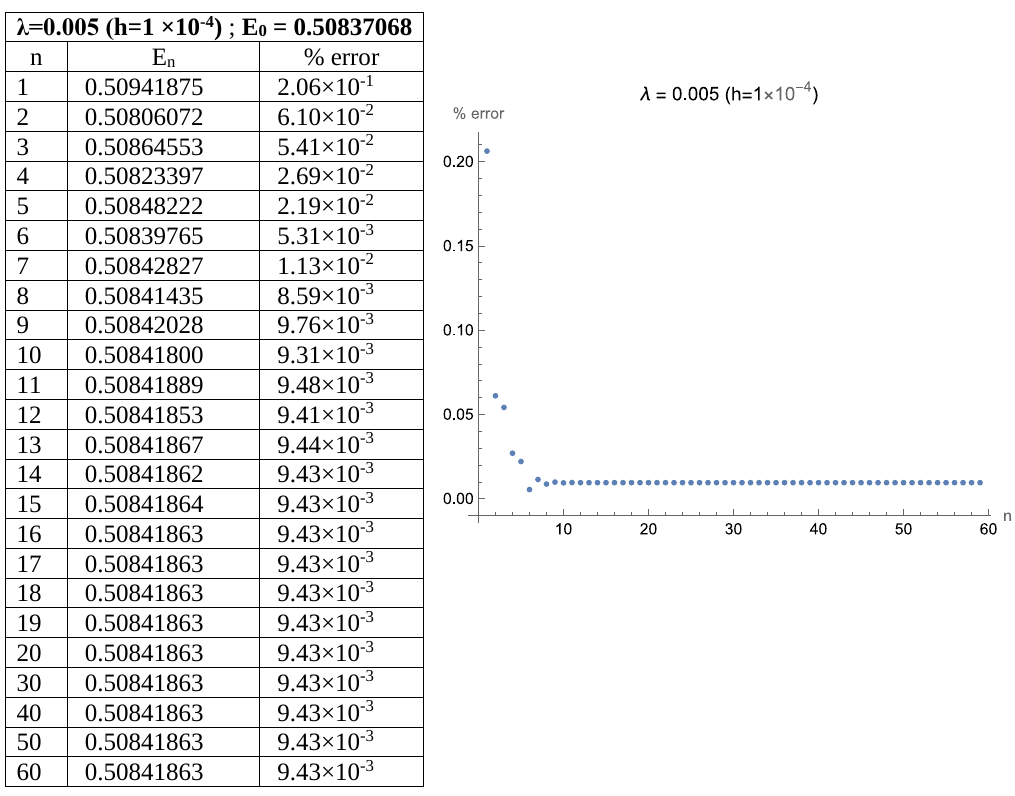}
		\caption{Results at intermediate coupling $\lambda=0.005$ with $h=1 \times 10^{-4}$. The parameter $h$ is smaller by a factor of $10$ compared to the previous case with $h=0.001$. The series now converges to the exact value to within less than $0.0095\%$ which is extremely low and basically negligible.}     
		\label{Inth2Sextic}
\end{figure}

\begin{figure}[t]
	\centering
		\includegraphics[scale=0.8]{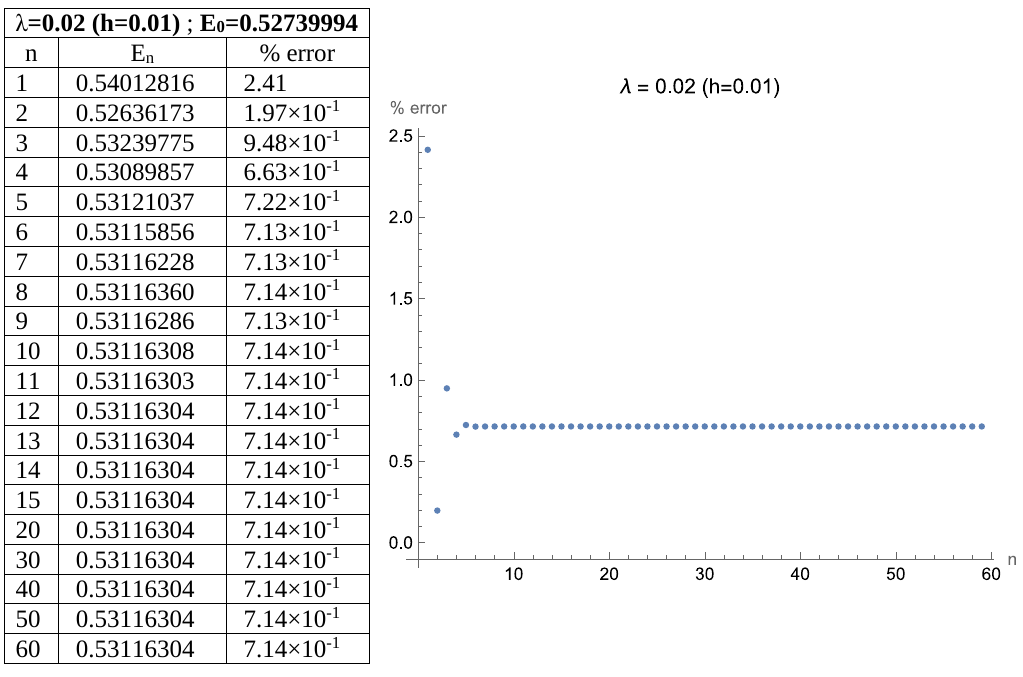}
		\caption{Results at strong coupling $\lambda=0.02$ for $h=0.01$. The series converges to the exact value to within $0.7 \%$ up to arbitrary large orders (shown here up to $n=60$). In scontrast the usual perturbative serie breaks down completely at strong coupling and diverges quickly from the start (see Fig. \ref{StrongSext}). This is an absolutely convergent series at strong coupling.}     
		\label{Strongh1Sextic}
\end{figure}

\begin{figure}[t]
	\centering
		\includegraphics[scale=0.65]{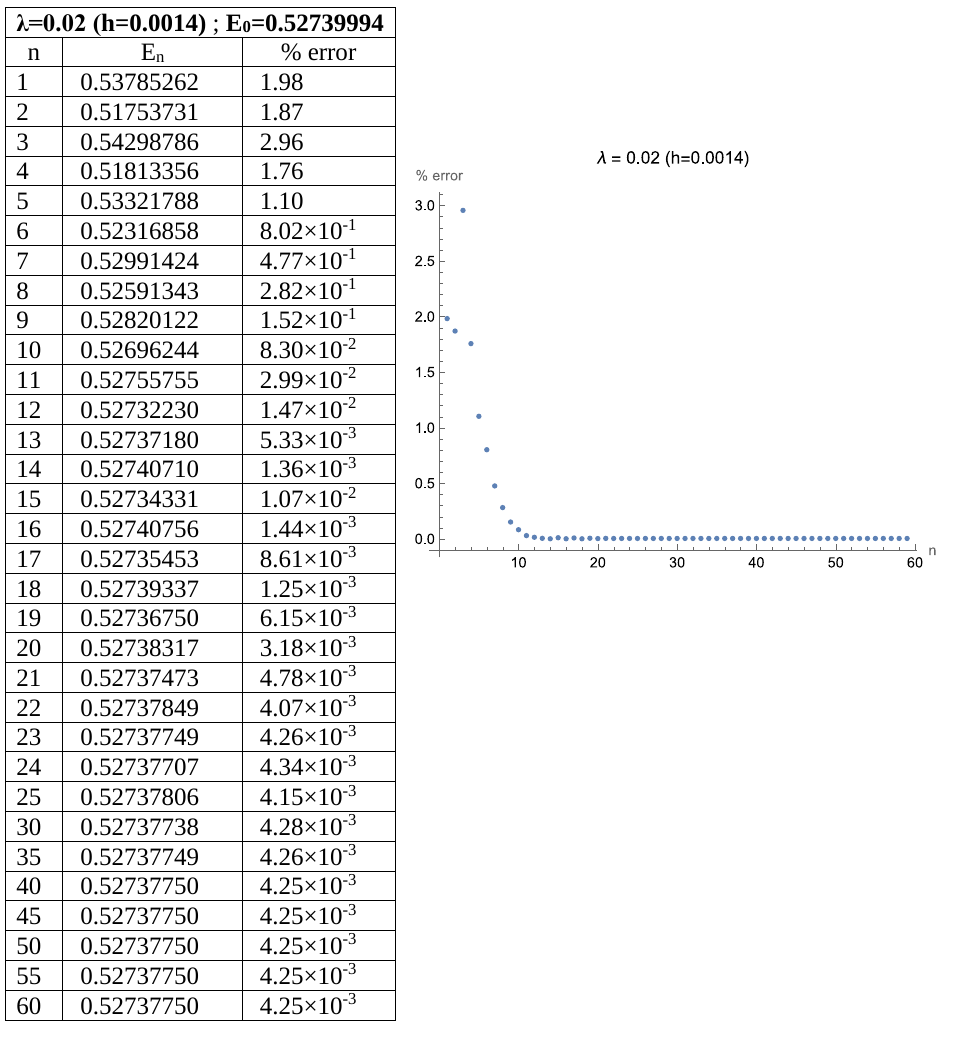}
		\caption{Results at strong coupling $\lambda=0.02$ for $h=0.0014$. The parameter $h$ is smaller by a factor of $\approx 7$ compared to the previous case of $h=0.01$. The series converges to the correct answer to within $0.004\%$ (compared to $0.7\%$ at $h=0.01$). Such a tiny negligible error is remarkable for a perturbative series at strong coupling. The is an absolutely convergent series (shown here up to $n=60$).}     
\label{Strongh2Sextic}
\end{figure}

Results are presented in figures \ref{Weakh1Sextic} to \ref{Strongh2Sextic}. Each figure contains a plot, a table of values and a description of the results for three different couplings: $\lambda=0.0005$ (weak coupling), $\lambda=0.005$ (intermediate coupling), and $\lambda=0.02$ (strong coupling). At each coupling, we show results for two values of the parameter $h$ where the second value is the lowest (i.e. going lower did not change the results within the given accuracy). We calculate the $\%$ error between $E_n$ and the exact value $E_0$ at each order $n$ (both $E_n$ and $E_0$ are quoted to eight decimal places). For each coupling at a given $h$, we plot the $\%$ error versus the order $n$. For a given coupling, the series converges closer to the exact energy for the smaller of the two values of $h$ as it should. At the two values of $h$ for each coupling, the series is an absolutely convergent series. The series matches the correct value at all couplings to within less than $0.8\%$ for all cases. The most important aspect of the results is that we obtained a convergent perturbative series in powers of the coupling at strong coupling to within a completely negligible error (less than $0.005\%$). The usual perturbative series for the sextic case diverges more strongly than the quartic case. So the results obtained here show that the convergent perturbative series based on finite path integral limits is very robust and in particular works remarkably well at strong coupling.    




\section{Conclusion}

In this work, we showed that if finite path integral limits are used instead of infinite limits, one obtains a convergent perturbative series in powers of the coupling instead of an asymptotic series. The convergence of the series implies that it can be used to make reliable calculations at strong coupling. We applied this to basic integrals containing quadratic and quartic terms in section 2 and to the ground state energy of the quartic and sextic anharmonic oscillator at strong, intermediate and weak coupling in sections 4 and 5 respectively. The basic integrals yield exact analytical expressions. The usual perturbative series in powers of the coupling $\lambda$ yields an asymptotic series but with finite integral limits they yield a convergent series for any coupling $\lambda$ and moreover, we are able to reproduce the analytical expressions. This was true even for the case with parameter $a<0$ where the asymptotic series was not Borel summable. For the anharmonic oscillator, the series expansion for the ground state energy in powers of the coupling is well-known to yield an asymptotic series \cite{Wu1,Wu2,Bender1,Bender2,Marino, Strocchi}. If one uses Schr\"odinger's equation to obtain the energy, finite integral limits are equivalent to placing infinite walls on both sides of the potential (i.e. at positions $-L$ and $L$ where $L>0$). We extracted the coefficients that enter the perturbative series from a new recursion relation that we derived for this specific scenario for both the quartic and sextic cases. The coefficients are a function of a positive parameter $h$ that encode the separation between the walls. The series for the energy for any given $h>0$ converges and approaches the ground state energy of the original potential (no walls) formally as $h\to 0$ ($h\ne 0$). For sufficiently small $h$, the series converged to and matched the exact (correct) values of the ground state energy to within tiny $\%$ errors for all coupling strengths including weak, intermediate and strong. The results for the quartic case are presented in figures \ref{Lambda002B}-\ref{Lambda22BB} and for the sextic case in figures \ref{Weakh1Sextic}-\ref{Strongh2Sextic}. At strong coupling the error is less than $0.1\%$ which is a remarkable result in light of how badly the usual perturbative series at strong coupling diverges (see figures \ref{Lambda022B} and \ref{StrongSext} for the quartic and sextic cases respectively). Using a perturbative series with finite integral limits, we obtained an observable at strong coupling, the ground state energy, for a dynamical quantum mechanical system. It would be of interest to apply the procedure in the future to the double-well in quantum mechanics as this potential is qualitatively different (e.g. it is not Borel summable). The main future goal would be to apply the procedure to realistic QFTs like QED and QCD at strong coupling. This should prove to be more challenging technically but in principle, the procedure should work. 
   
Dyson gave a famous argument as to why we should expect our expansions in powers of the coupling to yield an asymptotic series \cite{Dyson}. In the context of the anharmonic oscillator, the argument is simply that if the series about $\lambda=0$ were convergent at positive coupling $\lambda$, then it would also be convergent at negative coupling $-\lambda$ for sufficiently small absolute values. But $-\lambda$ leads to an unstable potential where tunneling occurs. It follows that such an unstable physical situation cannot yield a convergent series and therefore the same should hold true for the positive $\lambda$ case. How do finite integral limits avoid Dyson's argument?  Recall that in the context of Schr\"odinger's equation, finite integral limits meant that we had to place infinite walls on both sides of the potential. If the coupling is negative, the infinite walls on both sides prevent tunneling from occurring. There is no longer an unstable vacuum and Dyson's argument no longer applies. 

Finite integral limits originate from understanding why our usual perturbative series diverge even though the exponential in the integrand of a path integral has a series expansion with an infinite radius of convergence. This understanding consists of two parts. First, in the original path integral, the integrand contains an exponential to the power of an action containing an interaction part and a quadratic part. The path integral limits go up to infinity, and in the infinite limit the interaction part dominates over the quadratic part because it contains a product of more than two fields (e.g. in $\lambda \,\phi^4$ theory, $\phi^4>>\phi^2$ as $\phi \to \infty$). So this feature needs to be reproduced by the perturbative series. Secondly, the perturbative series is based on the expansion of the interaction part. The series expansion for an exponential to the power of the interaction cannot represent the exponential at the infinite limit of the integral (as $\phi\to \infty$) (e.g. if $e^{-\phi^4}$ appears in the original path integral, it is not equal to its series expansion in the asymptotic limit $\phi \to \infty$. The exponential is zero while the series up to any order $N$ diverges as $\phi^{4\,N}$ in that limit). So an important feature of the original path integral is not reproduced by the perturbative series and this shows up as a divergence. The issue is resolved if the terms in the perturbative series stem from integrals with finite limits (e.g. for finite arbitrary $\phi$, $e^{-\phi^4}$ is equal to its series expansion to any desired level of accuracy depending on how many terms one adds). Then one obtains an absolutely convergent series instead of an asymptotic one. This means that if a physical system is well defined by its path integral, its perturbative series based on finite integral limits will be an accurate representation of the physical system for sufficiently large limits.    

The basic integral with $a<0$ studied in section 2.1.2 yields the exact analytical expression given by \reff{I-} but its perturbative series \reff{Series1-} is an asymptotic series which is not Borel summable. With finite integral limits, we obtained the convergent series \reff{Series2-} expressed in powers of the coupling and the incomplete gamma function $\gamma(s,x)$ defined by \reff{gamma}. By using the series representation \reff{Seriesgamma} of the gamma function we recovered the original analytical expression \reff{I-} in the appropriate limit. So this example shows clearly that finite integral limits can work in cases where the asymptotic series is not Borel summable. It remains an open question whether this holds true in all non-Borel summable cases such as those stemming from renormalons in QED and QCD that we mentioned in the introduction. 

It is worthwhile to recall that finite integral limits are required only for the usual perturbative asymptotic series where the interaction term is expanded in powers of the coupling. As previously mentioned, a different series in inverse powers of the coupling was developed where instead of expanding the interaction term, one expanded the quadratic term  \cite{Edery}. This also yields an absolutely convergent series but no finite integral limits are required.

\end{document}